\documentclass[12pt]{article} 

\usepackage{latexsym,amssymb,amsfonts}
\usepackage[dvips]{graphicx}
\usepackage{amsmath}
\usepackage{graphicx}
\newcommand{\ba}{\begin{array}}
\newcommand{\ea}{\end{array}}
\newcommand{\be}{\begin{equation}}
\newcommand{\ee}{\end{equation}}
\newcommand{\bea}{\begin{eqnarray}}
\newcommand{\eea}{\end{eqnarray}}
\newcommand{\beaa}{\begin{eqnarray*}}
\newcommand{\eeaa}{\end{eqnarray*}}

\newcommand{\basa}{\begin{assumption}}
\newcommand{\easa}{\end{assumption}}

\newcommand{\bas}{\begin{assum}}
\newcommand{\eas}{\end{assum}}

\begin{document}
\title{Transport on river networks: A dynamical approach}

\author{Ilya Zaliapin \thanks{Department of Mathematics and Statistics,
University of Nevada, Reno, USA.
E-mail: zal@unr.edu.}, 
Efi Foufoula-Georgiou \thanks{St. Anthony Falls Laboratory and Department of 
Civil Engineering, University of Minnesota -- Twin Cities, Minneapolis, Minnesota, USA.
E-mail: efi@umn.edu.}, and 
Michael Ghil \thanks{D\'{e}partement Terre-Atmosph\`{e}re-Oc\'ean and
Laboratoire de M\'{e}t\'{e}orologie Dynamique,
Ecole Normale Sup\'{e}rieure, Paris, FRANCE and
Department of Atmospheric and Oceanic Sciences and
Institute of Geophysics and Planetary Physics,
University of California Los Angeles, USA.
E-mail: ghil@atmos.ucla.edu.}}
\maketitle

\begin{abstract}
This study is motivated by problems related to environmental transport on river networks.
We establish statistical properties of a flow along a
directed branching network and suggest its compact parameterization.
The downstream network transport is treated as a particular case of nearest-neighbor
hierarchical aggregation with respect to the metric induced by the branching structure
of the river network.
We describe the static geometric structure of a drainage network
by a tree, referred to as the {\it static tree}, and introduce an associated
{\it dynamic tree} that describes the transport along the static tree.
It is well known that the static branching structure of river networks can be described by
{\it self-similar trees} (SSTs); we demonstrate that the corresponding dynamic trees are
also self-similar.
We report an unexpected {\it phase transition} in the dynamics of three river
networks, one from California and two from Italy, demonstrate the universal features
of this transition, and seek to interpret it in hydrological terms.
\end{abstract}

\section{Introduction}
The topology of river networks has been extensively studied over the past decades
and stream ordering schemes, as well as statistical self-similarity concepts, have
been explored to a considerable extent
[see
{\it Horton}, 1945;
{\it Strahler}, 1957;
{\it Shreve}, 1966;
{\it Tokunaga}, 1978;
{\it Mandelbrot}, 1983;
{\it La Barbera and Rosso}, 1987;
{\it Marani et al.}, 1991;
{\it Rodriguez-Iturbe et al.}, 1992;
{\it Peckham}, 1995;
{\it Badii and Politi}, 1997;
{\it Rodriguez-Iturbe and Rinaldo}, 1997;
{\it Turcotte}, 1997;
{\it Sposito}, 1998;
{\it Peckham and Gupta}, 1999;
{\it Pelletier and Turcotte}, 2000;
{\it Burd et al.,} 2000;
{\it Dodds and Rothman}, 2000;
{\it da Costa et al.}, 2002;
and references therein]

What has been less studied, however, is how the static topology of a river network
affects and is affected by the dynamical processes operating over this network.
For example, consider a directed tree that represents a river network, and assume we
are interested in the mixing of water, solutes and sediments as they move downstream
in reaches of variable lengths and merge at junctions of the river network.
One might want then to attach a ``metric'' to the nodes of this tree, such as the
distance to the nearest source, and consider the notion of a
{\it dynamic tree}, superimposed on the template of the underlying {\it static tree}.

%What has been less studied, however, is how the static topology of a river network
%is affected by the dynamical processes operating over this network.
%For example, consider a directed tree that represents a river network.
%When one attaches a ''measure'' to the nodes of this tree, such as the distance to
%the nearest source (which can be used as a proxy for the time a pollutant takes to
%travel from the source to a given point in the network), intensity of rainfall or
%other fluxes which might change dynamically and heterogeneously along the static
%tree network, then one is confronted with the idea of considering the notion of a
%{\it dynamic tree} superimposed on the template of the underlying {\it static tree}.
This dynamic tree is likely to have a different hierarchy and topology than the static one.
Depending on the  dynamics, for example, some of the static-tree
branches might be completely cut off, either due to a blockage that prevents
transport along these branches or due to the absence of conditions to generate
sediment or nutrient for downstream transport.
In this case, the dynamic tree will have a different hierarchy than the static one,
and this difference might affect the scaling of fluxes that participate in defining
the envirodynamics on the network of interest.
In general, a static tree of a given Horton-Strahler order could become a dynamic
tree of a lesser or higher order, depending on the superimposed dynamics.

The purpose of this paper is to study the dynamic topology of
directed trees, starting with several simple cases, first synthetic and then realistic.
The direction we use is obviously ``downstream," {\it i.e.}, from the leaves to the root of the tree.
We focus on a dynamic hierarchy built on the concept of ``connectivity'':
once two streams are connected, they both influence the downstream dynamics.
One can thus imagine that two order-1 streams of different lengths, $l_1$ and
$l_2$, merge at a node but do not automatically give rise to an order-2 stream,
as would be the case in the standard Horton-Strahler ordering scheme; instead, we
keep track of length and the assigned order becomes 2 only when the running
index of length becomes  $\max(l_1,l_2)$.

Alternatively, one might keep track of time, rather than length: the two are equivalent
if the flow velocity is constant along all the branches, which we will assume in the
present paper, for simplicity's sake.
In other words, a dynamical node of order 2 is created only when the fluxes from
both order-1 streams do reach the connecting node.
Such considerations will result in a different ordering of the dynamic tree compared to
the static one. Moreover, the newly created dynamic tree will be {\it time-oriented},
a property that is absent in conventional static trees.

We approach the problem of hierarchical dynamics of river networks using general concepts
of {\it hierarchical aggregation}, which studies how multiple individual particles
(molecules, species, individuals, {\it etc.}) merge (aggregate, collide) with each other
to form clusters in different physical, chemical, biological, or sociological settings.
A major role in such studies is played by the notion of {\it cluster dynamics}.
This concept refers to the situation when a system that contains an infinite number of
interacting particles can be decomposed into {\it finite} clusters that move independently
of each other for some random interval of time.
After this time, the particle interactions give rise to infinite-range correlations,
although the system can be decomposed into another set of finite independent clusters,
and so on.

In the 1970s, Ya. G. Sinai developed a self-consistent mathematical formalism and
proved the existence of cluster dynamics for some particle systems in statistical mechanics
[{\it Sinai,} 1973, 1974].
The ideas of cluster dynamics have been applied to plasma physics, economics,
and the study of precursory patterns for extreme events in geophysics
[{\it Rotwain et al.}, 1997; {\it Molchan  et al.}, 1990; {\it Keilis-Borok and Soloviev}, 2003].
Recently, {\it Gabrielov  et al.} [2008] evaluated numerically the cluster
dynamics of elastic billiards, leading to the detection of what appear
to be the first genuine {\it phase transitions} and {\it scaling phenomena} that develop
in time, rather than with respect to a control parameter, such as temperature $T$ or density;
{\it i.e.}, a transition occurs and scaling develops as time $t$ evolves toward a critical
value $t^*$, rather than as the parameter $T$ crosses a critical value $T^*$.

In this paper, we adapt the concept of cluster dynamics to environmental
transport on river networks.
Notably, we obtain a remarkably similar, and equally unexpected, phase
transition in the cluster dynamics of river networks and attempt to interpret it in this context.
We also study the statistical properties of the dynamic trees introduced herein.
It is well known that the static branching structure of river networks can be described by
{\it self-similar trees} (SSTs); we demonstrate, using three actual river basins, that
the corresponding dynamic trees are also self-similar.

This paper is structured as follows.
We review in Section ~\ref{def} the terms and concepts relevant to the
hierarchical analysis of branching structures,
including the Horton-Strahler and Tokunaga branching taxonomies.
Section~\ref{DT} introduces the concept of a {\it dynamic tree} that is
associated with a given {\it static tree}, by using two examples from river transport.
Section~\ref{SH} describes two types of static trees analyzed in this study.
The first type reflects the well-formed ``river network'' of a basin.
The second type reflects the ``unchannelized drainage network''; this network is composed of
drainage paths that are not permanent channels but are perpendicular to the topographic
contour lines and follow the steepest downstream gradient. In other words, this drainage
network is formed by paths of ``zero-order'' basins or hillslopes.
Hierarchical aggregation is described in greater depth, and with additional examples
from several fields, in Section ~\ref{SH}, along with an abstract metric space setup.
Three actual river networks, from California and Italy, are analyzed in Section ~\ref{rivers}.
A summary and discussion follow in Section ~\ref{discussion}.

\section{Main concepts and definitions}
\label{def}
This section introduces the main concepts used in the analysis of
branching structures, along with their definitions and illustrative
examples.

\subsection{Trees}
\begin{figure}[p] %[p] [t]
\centering\includegraphics[width=.8\textwidth]{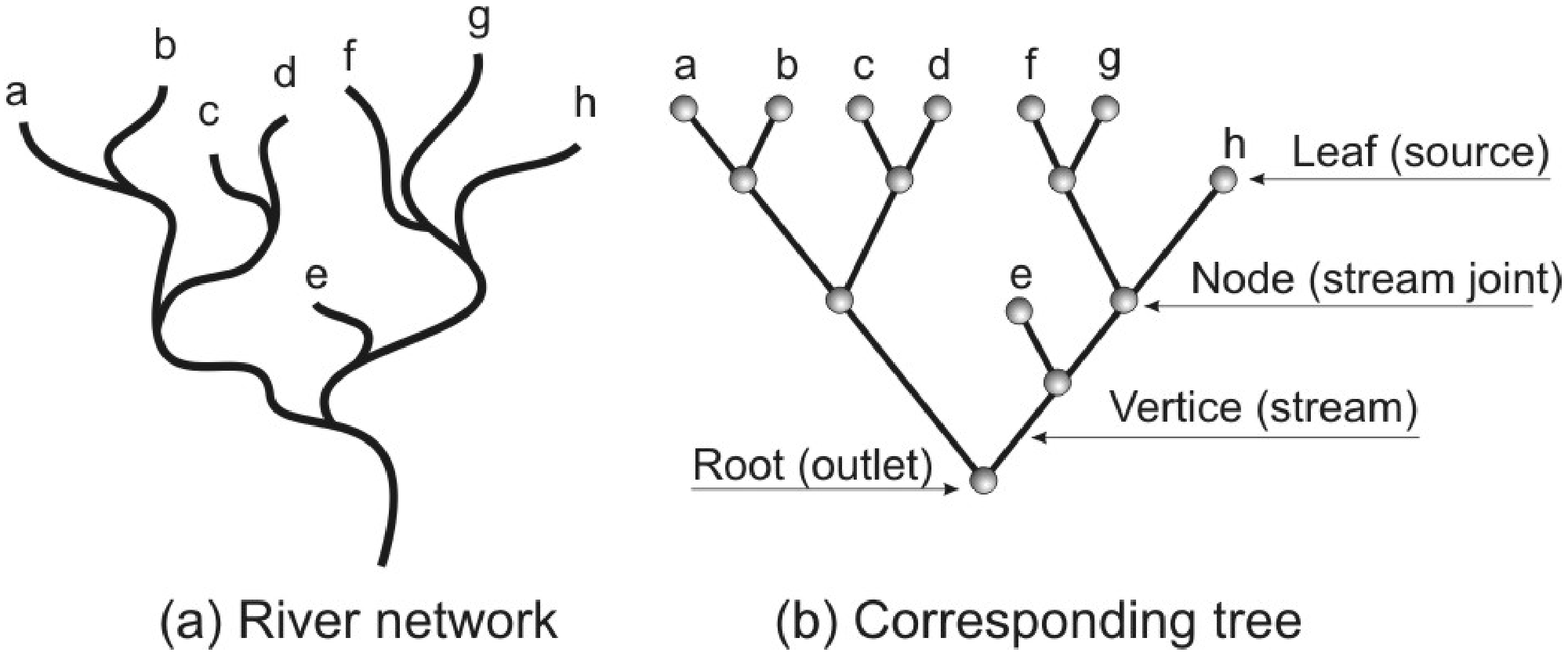}
\caption[Tree representation of a river network.]
{Tree representation of a river network: (a) hypothetical river network;
and (b) its representation by a binary tree.
The network sources and the respective tree leaves are marked by the
same letters in both panels.
The figure also illustrates the terminology used in our river transport study. }
\label{scheme}
\end{figure}

\indent
\label{tree}
A {\it tree} $\mathbb{T}$ is a set of {\it nodes} connected by {\it vertices}
(also called {\it edges} or {\it links})
in such a way that there are no loops, {\it i.e.} there are no closed
paths formed by distinct edges (see Fig.~\ref{scheme}).
A {\it rooted tree} has one special node designated as a {\it root}.
In a rooted tree each connected pair of nodes has a parent-child relationship,
with the parent being the element that is closest to the root [{\it Athreya and Ney}, 1972].
The nodes with no children are called {\it leaves}.
The {\it depth} $d_i$ of a node $i$ in a rooted tree is defined as the number of
edges between this node and the root.
The depth $D$ of a tree is the maximum of the depths $d_l$ over all the leaves $l$.

In this study we will work with {\it binary trees}.
In a binary rooted tree, each node may have either two or no children.
This means that each internal node $i$ (every node except for the root and the leaves)
is connected to three other nodes: one is a parent of $i$, and the other two
are its children.
The root is only connected to two children, and each leaf is connected to a single parent.
A {\it complete binary tree} is a rooted tree such that all its leaves have the same depth.
Our interest for binary trees is motivated by the observation that
many natural phenomena exhibit binary branching.
For example, in river networks, it is unlikely for three or more streams to
merge at exactly the same point, while in gas dynamics it is unlikely that more than two
molecules will collide at the same time.

In our study of river transport, the tree $\mathbb{T}$ will represent a drainage network;
see Fig.~\ref{scheme}.
Hence the nodes correspond to the merging points of streams and vertices
to the stream segments between these points, while
the network's sources are the leaves, and the outlet is the root of the tree.

\subsection{Branching-order taxonomies}
\indent
\label{branch}
	In many applications, there is a need to order the nodes according to
their importance in forming the entire hierarchy; this importance often corresponds also
to relative size.
For instance, in a botanical tree the leaves are the most delicate, smallest
elements; the intermediate levels are formed by consecutively
wider branches, while the most heavy, robust element of the plant is its trunk.
Likewise, one naturally distinguishes in a river network between minor and
major tributaries, according to the amount of water that they are able to carry.

In a complete tree, the node ordering task is quite straightforward since a node's order
can be chosen to be inversely proportional to its depth: ``the deeper, the smaller''.
The problem, however, becomes more complicated when one deals with an incomplete tree;
in this case, the depth can no longer serve as a proxy for size, since the leaves, while being
the smallest elements, will often be assigned indices that are as large as those of much
heavier internal nodes.

{\it Horton} [1945] developed a convenient way to order hierarchically organized
river tributaries; this method was later refined by {\it Strahler} [1957] and further
expanded by {\it Tokunaga} [1978].
Currently, the so-called Horton-Strahler and Tokunaga ordering schemes are standard
tools of branching analysis.

\subsubsection{Horton-Strahler ordering}
\begin{figure}[p] %[p] [t]
%\centering\includegraphics[width=.45\textwidth]{figures/HS_ex.eps}
\centering\includegraphics[width=.9\textwidth]{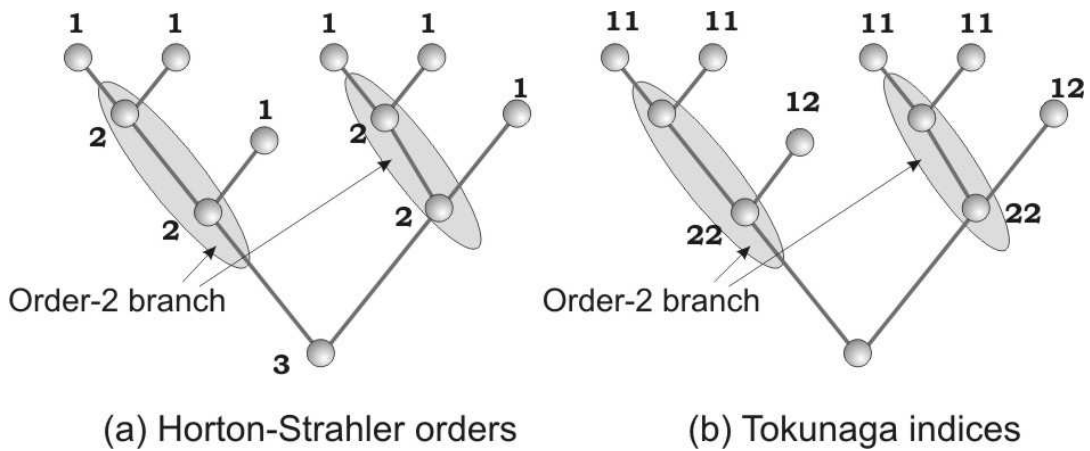}
\caption[Example of Horton-Strahler indexing]
{Example of (a) Horton-Strahler ordering, and (b) Tokunaga indexing.}
\label{fig_HST}
\end{figure}

\indent
\label{hs}
Each leaf in a binary rooted tree is assigned a Horton-Strahler (HS) {\it order}
$r({\rm leaf})=1$; see Fig.~\ref{fig_HST}a.
Each node $p$, which is the parent of nodes $c_1$ and $c_2$, is assigned a
Horton-Strahler order $r(p)$ according to the following rule
[{\it Horton}, 1945; {\it Strahler}, 1957; {\it Newman et al.,} 1997]:
\begin{equation}
r(p)=\left\{
\begin{array}{ll}
r(c_1)+1&{\rm if~}r(c_1)=r(c_2)\\
\max\left(r(c_1),r(c_2)\right)&{\rm if~}r(c_1)\ne r(c_2).
\end{array}
\right.
\label{HS}
\end{equation}

A {\it branch} is defined as a union of connected nodes with the same order.
We will denote by $N_r$ the total number of branches of order $r$.
Notice that each branch has {\it linear} structure: two children of the same
parent can not belong to the same branch.

In a tree with $n$ leaves, the longest branch can be formed by $(n-1)$ nodes;
this is the case when two leaves merge together to form an order-2 branch and
then all other leaves join this branch one by one.
We refer to this situation as {\it exhaustive branching}.
It is readily seen that each leaf is always an order-1 branch.
An order-2 branch is created by merging two leaves and can consist of more
that one node, depending on the leaves that join it; an order-2 branch that
consists of two nodes is highlighted in Fig.~\ref{fig_HST}a.
The order $\Omega$ of a tree is the maximal order of its branches (or nodes).

In a complete tree, each branch consists of a single node since the children of
an order-$r$ node always have the same order $(r-1)$.
In such a tree, the HS order is uniquely determined by the node depth $d$ via
$r=D-d+1$, where the tree depth is $D=\Omega$.

\subsubsection{Tokunaga indexing}
\indent
\label{tokunaga}
Tokunaga indexing [{\it Tokunaga}, 1978; {\it Peckham}, 1995; {\it Newman et al.}, 1997]
extends upon the Horton-Strahler orders; it is illustrated in Fig.~\ref{fig_HST}b.
This indexing focuses on incomplete trees by cataloging the merging points between
branches of different order.
A first-order branch that merges with a second-order branch is indexed by
``12'' and the total number of such branches is denoted by $N_{12}$.
A first-order branch that merges with a third-order branch is indexed by
``13'' and the total number of such branches is $N_{13}$, and so on.
In general, $N_{ij}$ for $j>i$ denotes the total number of order-$i$ branches that
join an order-$j$ branch.

The Tokunaga index $T_{ij}$ is the number of branches of order $i$ that merge
with a branch of order $j$, normalized by the total number of branches of order $j$;
in other words, $T_{ij}$ is the average number of branches of order $i<j$ per
branch of order $j$:
\begin{equation}
T_{ij}= \frac{N_{ij}}{N_{j}}.
\label{tok}
\end{equation}

Merging of branches of different orders is referred to as {\it side branching}.
It is easily seen that side branching is absent in a complete tree, and
``a tree with side branching'' is synonymous to ``an incomplete tree.''
For incomplete trees, the side-branching indices become increasingly important as they
help to define a tree's structure, possibly indicating properties that
are unique to specific classes of trees.

For consistency, we denote the total number of order-$i$ branches
that merge with other order-$i$ branches by $N_{ii}$ and notice that
in a complete binary tree $N_{ii}=2\,N_{i+1}$.
This allows us to formally introduce the additional Tokunaga indices:
\[T_{ii}=\frac{N_{ii}}{N_{i+1}}\equiv 2.\]
The set $\{T_{ij}: 1\le i,j \le \Omega\}$ of Tokunaga indices provides a
complete statistical description of the branching structure of an order-$\Omega$ tree.

\subsubsection{Other node statistics}
\label{stats}
\indent
We introduce here two node statistics relevant to our river transport study:
\begin{itemize}
\item{the {\it number of nodes} (or links) within a branch $i$ is denoted by $c_i$}; and
\item{the {\it magnitude} $m_i$ of a node $i$ is the number of leaves that descend from $i$;
in other words, the magnitude of a branch is the number of the sources upstream of it.}
\end{itemize}

Magnitude measures the complexity of the river structure upstream from a given branch.
We notice that each leaf (source) has unit magnitude, $m_{\rm leaf}=1$, and the
magnitude of a parental node $p$ is the sum of the magnitudes of its children
$c_1$ and $c_2$:
\begin{equation}
m_p=m_{c_1}+m_{c_2}.
\label{mass}
\end{equation}
Accordingly, a node or order $r$ has magnitude $m\ge 2^{r-1}$, with
equality being attained only for a complete binary tree.
The average number of nodes and average magnitude of an order-$r$ branch are denoted
by $C_r$ and $M_r$ respectively.

\subsection{Self-similar trees}
\indent
\label{sst}
The concept of {\it self-similarity} provides a powerful tool for describing and studying trees.
A self-similar tree (SST) is defined by the constraint
\be
T_{i,i+k} = T_{k} \;\; \rm{for} \;\; k=1,2,\dots.
\ee

E. Tokunaga was probably the first to study SSTs, and
considered an additional constraint on the branching indices
[{\it Tokunaga}, 1978]:
\be
\frac{T_{k+1}}{T_k}=c,\quad{\rm or~} \quad T_{k}=a\,c^{k-1}\;\;{\rm for}\;\;a,c> 0.
\label{TSS}
\ee
The SSTs that satisfy \eqref{TSS} are called {\it Tokunaga trees}.

\subsection{Horton laws}
\label{HL}
Empirically, the average values of branching statistics for the observed river
basins depend exponentially on the order $r$:
\begin{align}
\label{HLN}
N_r&=N_0\,R_B^{\Omega-r},\\
\label{HLM}
M_r&=R_M^{r-1},\\
C_r&=C_0\,R_C^{r}
\label{HLC}
\end{align}
for some positive constants $N_0$ and $C_0$.
Such relationships are called {\it Horton laws};
the bases $R_B, R_M$, and $R_C$ of the exponential relatonships are
called {\it stream ratios}.

{\it McConnell and Gupta} [2008] showed that the Horton
laws \eqref{HLN}, \eqref{HLM} hold asymptotically, {\it i.e.} for $r\to\infty$, in a
self-similar Tokunaga tree; they also proved that $R_B=R_M$.
Moreover, {\it Zaliapin} [2009] demonstrated the stream ratio inequality
\be
R_B=R_M<R_C,
\label{ratios}
\ee
that had been conjectured by {\it Peckham} [1995].
In addition, {\it Zaliapin} [2009] demonstrated that the Horton laws hold,
under some additional assumptions on the Tokunaga indices $T_k$,
for self-similar trees that do not necessarily satisfy the
Tokunaga condition \eqref{TSS}.

\section{Static vs. dynamic trees: Network envirodynamics}
\label{DT}
The topological structure of a river network is well described by
a tree, which we denote by $\mathbb{T}_S$ and call the {\it static tree}.
To describe the downstream transport on $\mathbb{T}_S$ we now introduce a
{\it dynamic tree} $\mathbb{T}_D$, which can be interpreted as follows.
Imagine that we inject a dye simultaneously into all the sources of our river network,
represented by the leaves of $\mathbb{T}_S$, and the dye starts propagating down the river,
from the sources to the outlet, with the same constant velocity along all the streams.
The tree $\mathbb{T}_D$ describes the time-dependent history of the mergings of
the colored streams.

Next, we consider two detailed examples that will clarify this important concept.
We restrict ourselves to the simplest case of constant velocity along
all the streams; taking this velocity to be unity, time and length scales can be interchanged.
An extension to spatially or temporally variable velocities is straightforward:
we shall see that the dynamic tree $\mathbb{T}_{\rm D}$ is completely determined by
the static tree $\mathbb{T}_{\rm S}$ and the set of time {\it delays} $\tau_i$ necessary
for the dye to propagate from a node $i$ to its parent.

\subsection{Synthetic example}
\label{syn_ex}
\begin{figure}[p] %[p] [t]
\centering\includegraphics[width=\textwidth]{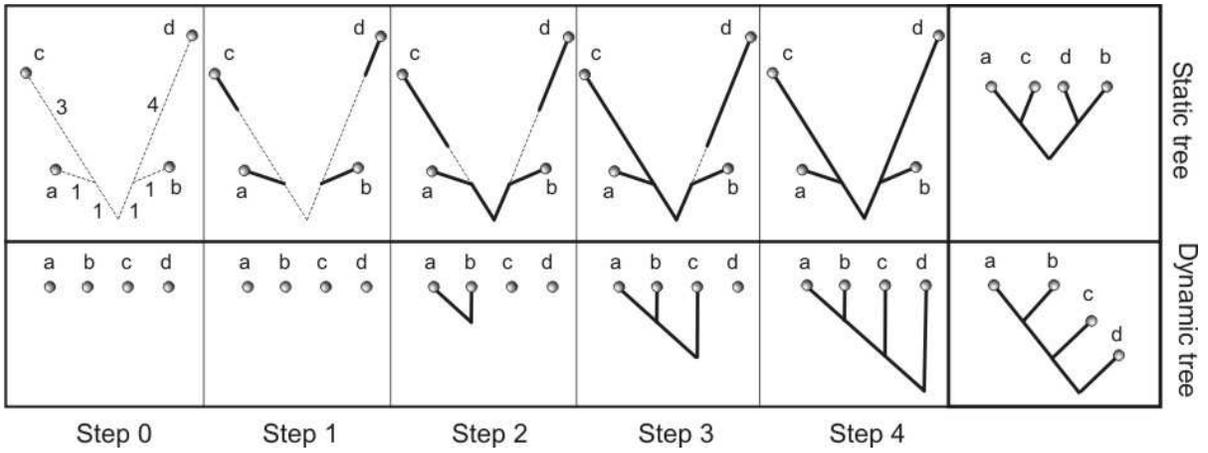}
\caption{Constructing a {\it dynamic} tree.
The initial static tree and the final dynamic tree are shown in the
rightmost pair of panels.
The dynamic tree reflects the propagation of a flux from leaves
to the root of the static tree at a constant velocity.
The top row of panels shows the static tree at different steps of this process;
for visual convenience we explicitly show the static tree's link lengths.
The bottom row shows the corresponding phases of the dynamic tree.
The top leftmost panel indicates the lengths of the links in the static tree;
each step in the figure takes one time unit, that is the flux propagates one unit
of length downstream.
See Section ~\ref{syn_ex} for details.}
\label{fig_dyn}
\end{figure}

Figure~\ref{fig_dyn} shows how to construct the dynamic tree for a basin
with four sources {\bf a, b, c,} and {\bf d}.
The static tree for this basin is a complete binary tree shown in the top right panel.
The same tree with the link lengths explicitly shown is placed in the
top row of panels; the top left panel indicates the values of these lengths.

The consecutive phases of construction of the dynamic tree are shown in the bottom
row of panels.
At step 0 (the leftmost top and bottom panels), all the links in the tree are
``empty'' (dashed lines) and the dye is injected into the sources {\bf a, b, c,} and {\bf d}.
Accordingly, we have four disconnected clusters of colored flux; they correspond
to four disconnected nodes in the lower left panel.
Each step in the figure is a snapshot of the process after a unit time interval;
recall that we only use constant velocity in this paper and, without loss of generality,
this velocity equals unity.

At step 1 the dye has propagated a unit length along each stream, which is depicted
by solid lines in the top panel.
Since all four streams are disconnected so far, the dynamic tree still consists of four
disconnected branches, each of which corresponds to a colored stream of unit length.
At step 2 the streams {\bf a} and {\bf b} merge.
This is reflected in the dynamic tree, where the nodes {\bf a} and {\bf b}
are now connected into a single cluster.
Notice that the leaves {\bf a} and {\bf b} are not directly connected in the static
tree; this connection reflects a special property of the dye's downstream propagation.

At step 3 stream {\bf c} reaches stream {\bf a}.
Since stream {\bf a} by that time is already merged with stream {\bf b},
we say that the stream {\bf c} merges with the cluster of {\bf a} and {\bf b};
this is reflected in the dynamic tree in the lower panel for this step.
Hence, at step 3 there exist two connected clusters of the colored flux: one cluster
is formed by the streams {\bf a, b,} and {\bf c}, while stream {\bf d}
alone forms the second cluster.
Finally, at step 4, all the colored fluxes merge together.
The conventional representation of both static and dynamic trees, which does not show
the link lengths, is given in the two rightmost panels.

This example shows that the dynamic tree $\mathbb{T}_D$ can be very different from the
corresponding static tree $\mathbb{T}_S$.
We notice in particular that in this example the static tree is a tree with no side
branching; it has the largest possible Horton-Strahler order, $\Omega=3$, for a tree
with four leaves.
At the same time, the dynamic tree exhibits exhaustive side-branching;
accordingly, it has the smallest possible order, $\Omega=2$, for a four-leaved tree.

\subsection{Realistic example}
\begin{figure}[p] %[p] [t]
\centering\includegraphics[width=.6\textwidth]{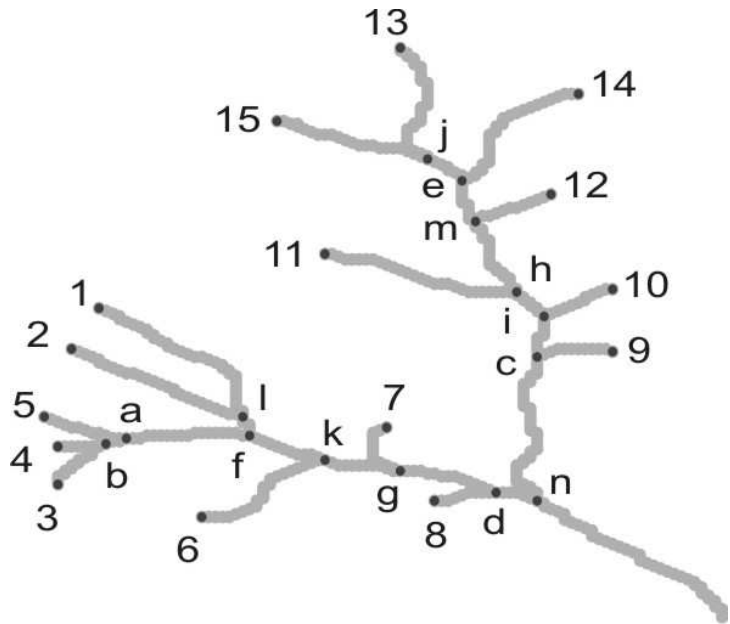}
\caption{Stream network for an order-3 subbasin of the Noyo river basin, Mendocino
county, California.
Sources are marked by letters, stream merging points by numbers.
The same marks are used in Figs.~\ref{Noyo_sub_2} and~\ref{Noyo_sub_3}
that show the static and dynamic trees for this stream.}
\label{Noyo_sub_1}
\end{figure}

\begin{figure}[p] %[p] [t]
\centering\includegraphics[width=\textwidth]{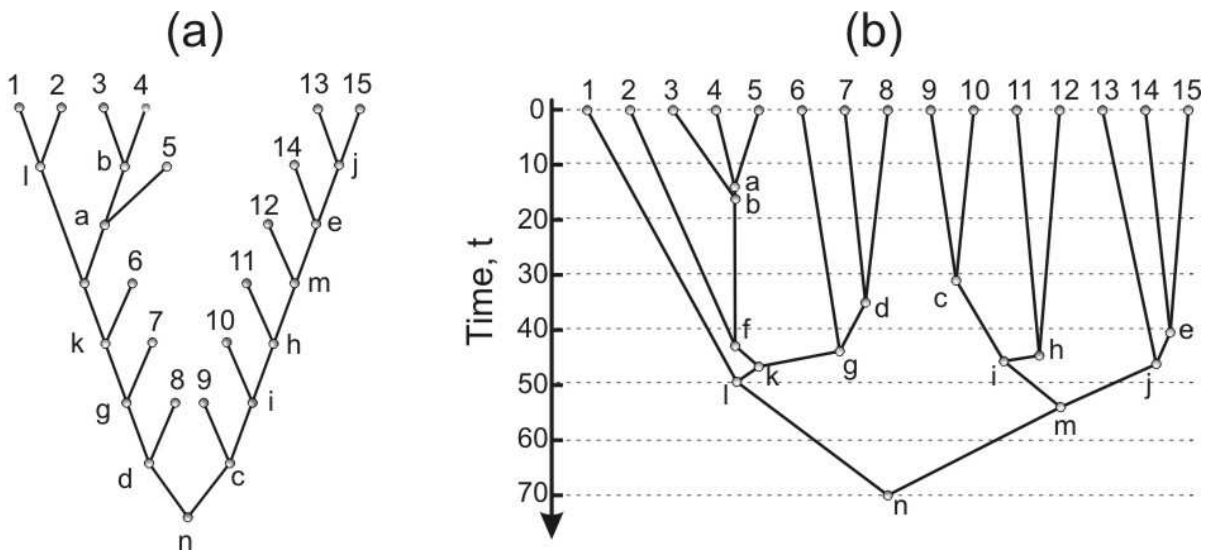}
\caption{Static and dynamic trees for the Noyo subbasin of Fig.~\ref{Noyo_sub_1}.
(a) Static tree $\mathbb{T}_{\rm S}$ and (b) dynamic tree $\mathbb{T}_{\rm D}$.
Letter and number marks are the same as in Fig.~\ref{Noyo_sub_1}.}
\label{Noyo_sub_2}
\end{figure}

Here we illustrate the dynamic tree for an order-3 subbasin of the Noyo basin; this basin
is located in Mendocino County, California, USA, and is described by {\it Sklar et al.} [2006].
The stream network for this subbasin is shown in Fig.~\ref{Noyo_sub_1}; its fifteen
sources are marked by numbers 1 to 15 and fourteen stream joints by letters {\bf a} to {\bf n}.
The static tree $\mathbb{T}_{\rm S}$ for this stream network is shown in Fig.~\ref{Noyo_sub_2}a;
it has the Horton-Strahler order $\Omega=3$.

The time-oriented dynamic tree $\mathbb{T}_D$ is shown in Fig.~\ref{Noyo_sub_2}b against
the time axis (on the ordinate); notice that time can also be interpreted as the distance traveled by
the dye from each source.
This interpretation has a direct connection to the metric properties of the basin and
we will use it in the subsequent analysis.
The order of the dynamic tree is $\Omega=4$.
The letter and number marks in Fig.~\ref{Noyo_sub_2} match those in Fig.~\ref{Noyo_sub_1}.

\begin{figure}[p] %[p] [t]
\centering\includegraphics[width=\textwidth]{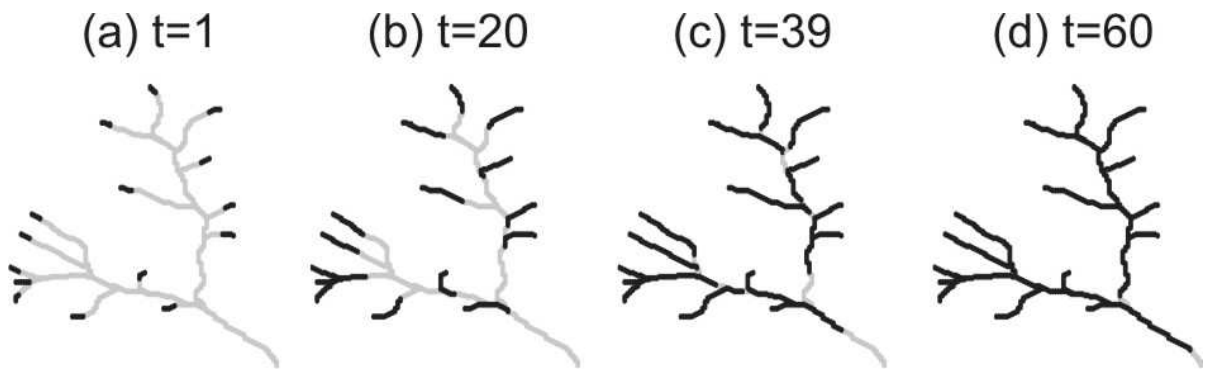}
\caption{Three snapshots of the evolution of the dynamic tree (heavy solid lines)
on the static tree (light solid lines) for the stream of Fig.~\ref{Noyo_sub_1}.
Letter and number marks are the same as in Fig.~\ref{Noyo_sub_1}.}
\label{Noyo_sub_3}
\end{figure}

Four snapshots of the dye propagation --- at times $t=1, 20, 39,$ and $60$ --- are shown
in Fig.~\ref{Noyo_sub_3}.
In this example, the dynamic tree shows a larger degree of side-branching compared
to the static tree; this larger degree is reflected in its larger HS order.
We shall see in other realistic examples, further below, that this seems to be the case for most
actual river networks.

\section{Stream vs. hillslope networks}
\label{SH}
In an actual landscape, channels are initiated when the area upstream suffices
to create a sustainable source of streamflow and this source imprints a permanent
channel on the terrain.
Although these channels are typically detectable by field observations, the extraction
of the channel initiation points, or ``channel heads,'' from Digital Elevation Models (DEMs)
has been a subject of intense study.
%({\it e.g.} ).
%Tarboton et al., 1991; Montgomery and Foufoula-Georgiou, 1993; Costa-
%Cabral and Burges, 1994; Giannoni et al., 2005; Hancock and Evans, 2006]
%})
Most commonly, channels are assumed to be initiated when the upstream area, or area
times a typical slope, exceed a given threshold; the parameters of such relationships are
field-calibrated. More recently, the availability of high-resolution, 1-m elevation data from
LIght Detection and Ranging (LIDAR) instrumentation has initiated a new generation
of methodologies for the automatic detection of channels as ``edges'' or ``features'' in
the terrain [{\it e.g.}, {\it Lashermes et al.}, 2007; {\it Passalaqua et al}, 2009].

The channelized  paths, {\it i.e.} the branches of the river network, are not the only parts
of the basin by which water or other fluxes --- {\it e.g.}, sediments, nutrients, or pollutants ---
are transported downstream.
The unchannelized part of the basin, often called {\it zero-order basins} or
{\it hillslopes}, is drained by pathways that have their own topology.
In this work, we extract (i) {\it stream networks} from DEMs by using a critical threshold
area $A_c$, and (ii) {\it hillslope networks} by assuming that $A_c$
is as small as the DEM resolution.

Clearly, each stream network is a part of the corresponding hillslope network.
For a generic river basin, though, the total length of channelized paths is
much smaller than the total length of unchannelized paths. In the present study,
we assume that stream networks reflect the properties of channelized paths,
while hillslope networks reflect the properties of unchannelized paths.
The study of unchannelized-path topology below will show that it is
quite different from the topology of the channelized paths.

\begin{figure}[p] %[p] [t]
\centering\includegraphics[width=.9\textwidth]{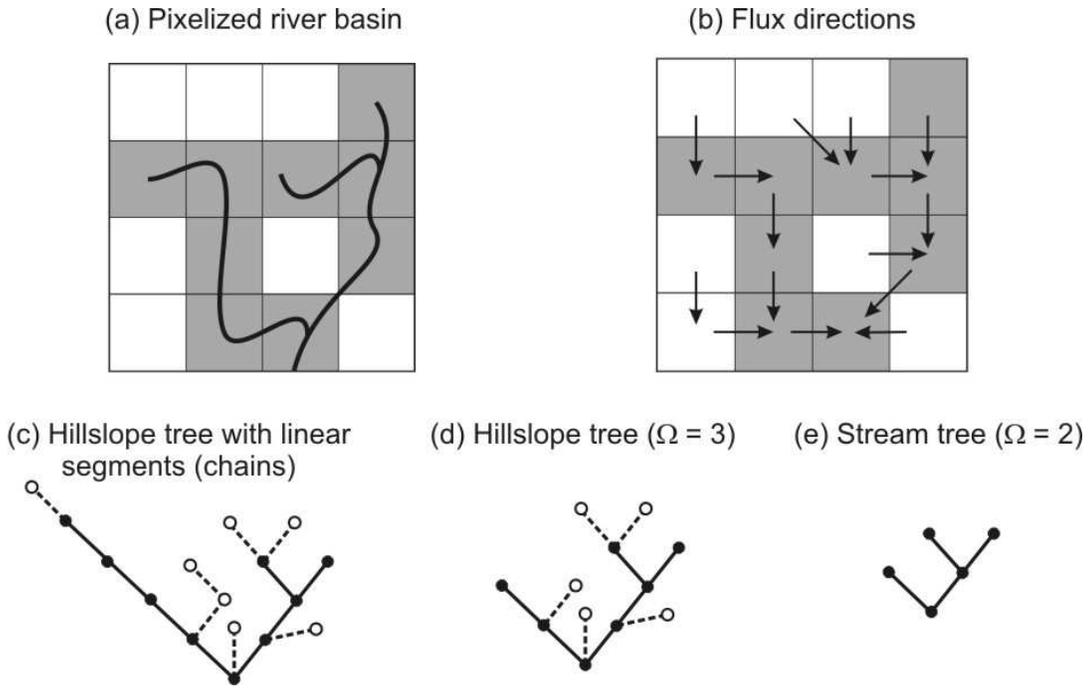}
\caption{Construction of a {\it static} tree that represents the topology of
hillslope (unchannelized) and stream (channelized) drainage paths.
(a) Pixelized river basin; the shaded pixels (cells) correspond
to the stream location (solid line), the white pixels -- to the valleys or
hillslopes.
(b) Flux direction obtained from the elevation data.
(c) Tree that describes the drainage topology: solid nodes and links correspond
to the stream pixels and stream flow; open nodes and dashed links -- to the hillslope
pixels and hillslope flow.
Notice that this tree contains several purely linear segments, with no branching.
(d) The same tree, from which the linear segments have been removed: it describes
the topology of both hillslope paths and stream paths, and is referred to as the {\it hillslope tree}.
(e) The subset of the tree in panel (d) that describes the topology of the stream paths
only; this tree is referred to as the {\it stream tree}.}
\label{fig_dir}
\end{figure}

Construction of stream and hillslope static trees is illustrated
in Fig.~\ref{fig_dir}.
Figure~\ref{fig_dir}a shows a small part of a river basin;
it is divided into 16 square regions called {\it pixels}.
The well-defined streams occupy some of the pixels (shaded squares), the rest
of the pixels (white squares) represent {\it hillslopes}, {\it i.e.} unchannelized parts
of the basin.

The elevation data can be used to figure out the flux direction from each pixel,
whether stream or hillslope; this direction is depicted by arrows in Fig.~\ref{fig_dir}b.
We assume that there is a unique flux direction away from each pixel;
at the same time, fluxes can reach a given pixel from more than one
other pixel.
This property allows one to represent the directional information by a tree,
which is shown in Fig.~\ref{fig_dir}c.
Solid nodes and solid lines in the figure represent the stream pixels and
the stream flow respectively, while open nodes and dashed lines represent
the hillslope nodes and hillslope flow.

The final step in creating the static tree of this subbasin is to remove the linear segments (chains),
that is to remove the nodes with only two connections (except the tree root).
The resulting static hillslope tree is shown in panel (d).
The static stream tree is obtained from the hillslope tree by removing
the dashed links that represent unchannelized paths, and removing
the remaining chains; the stream tree for our example is shown in panel (e).

\section{The dynamics of hierarchical aggregation}
\label{HA}
The consecutive merging of river streams discussed in the previous section
is a special case of a general phenomenon of {\it hierarchical aggregation}.
This phenomenon is also called {\it inverse cascading}, and it can be described as follows.

Consider a process that starts at time $t=0$ with $N$ individual {\it particles},
which can be considered as {\it clusters} of unit mass.
As time evolves, the clusters start to merge with one another, according
to a set of suitable rules, thus forming consecutively larger clusters.
We assume that only two clusters can merge at the same time; thus
after each merging the number of clusters decreases by one.
The process continues until all particles have been merged into a single
cluster of mass $N$.
The evolution of the above process can be described by a time-oriented binary tree,
whose leaves correspond to the initial particles, the root to the final cluster of $N$
particles, and each internal node to the merging of a particular pair of clusters.

\subsection{Examples}
\label{exs}
Among the many instances of the above general aggregation scheme,
we mention here the following four.

\paragraph{Percolation.} In the {\it site percolation} process on an $L\times L$
lattice, the initial $N=L^2$ particles correspond to the sites of the lattice, while
clusters correspond to connected patches of occupied sites that are formed
during the percolation process [{\it Zaliapin et al.}, 2005]. In fact, the same scheme can be
applied to bond percolation, as well as to percolation on grids in higher dimensions.

\paragraph{Billiards.} {\it Elastic billiard} on a rectangular table
can be used to model {\it gas dynamics} in two dimensions (2-D). Here the initial particles
are the $N$ billiard balls (gas molecules) at time $t=0$.
Each of the balls is assigned an initial position and velocity.
The clusters at time $\Delta$ are formed by balls that have collided during the
time interval $[0,\Delta]$ [{\it Gabrielov et al.}, 2008].
Formally, two balls are called $\Delta$-{\it neighbors} if they collided during the
time interval $[0,\,\Delta]$.
Each connected component of this neighbor relation is called a $\Delta$-{\it cluster}.
Notice that within an arbitrary $\Delta$-cluster each ball has collided with at least
one other ball during the time interval $[0,\,\Delta]$.
In other words, a $\Delta$-cluster is a group of balls that have affected each
other's dynamics during the time interval of duration $\Delta$.
The mass of each cluster is simply the total number of balls within that cluster.
Upon many collisions of the balls, the whole system will be composed of clusters
of different sizes.
As time evolves, the number of clusters will decrease and their mass increase.

The same scheme can be applied to a system of particles that interact according to
some potential $U({\bf x})$.
{\it Bogolyubov} [1960] suggested that when the interaction of particles
is short-ranged, the system can be decomposed into finite clusters so that during
some random interval of time, each cluster moves independently of other clusters as
a finite-dimensional dynamical system.
After this time interval, the system can be decomposed again into other dynamically
independent clusters and so on.
This type of dynamics is called {\it cluster dynamics} and {\it Sinai} [1974] showed
analytically that it exists in a one-dimensional (1-D) system of statistical mechanics.
Numerical results of {\it Gabrielov et al.} [2008] describe the presence and various
properties of cluster dynamics in a 2-D system of hard balls.

\paragraph{Phylogenetic trees.}
Probably the best-known application of hierarchical aggregation is in constructing
phylogenetic trees that describe the evolutionary relationships among biological
species [{\it Maher}, 2002]. Here, a node corresponds to a set of species.
Two species are connected if they have a direct common ancestor; the link length from
a species to its direct ancestor equals the time it took to develop the descendant species
from that ancestor.

\paragraph{River transport.}
The example of interest to us here is the downstream transport along a river network.
In this case, the initial particles are the
environmental fluxes at the sources of the network, and clusters are formed by consecutive
merging of the streams down the river path.
That is, new clusters are formed when fluxes from upstream merge
at the stream junctions.
This scheme of describing dynamics along a static tree was considered in detail in
Section ~\ref{DT}, albeit without referring to hierarchical aggregation.

\subsection{General set-up}
\label{gen}
Hierarchical aggregation can be described in great generality by using the framework
of nearest-neighbor clustering in a metric space.
Specifically, consider a set $\mathbb{S}$ with distance $d(a,b)$
for $a,b\in\mathbb{S}$; the elements of the set will be called {\it points}.
The distance $d(A,B)$ between two subsets of points $A=\{a_i\}_{i=1,\dots,N_A}$ and
$B=\{b_i\}_{i=1,\dots,N_B}$ from $\mathbb{S}$ is defined as the shortest distance
between the elements of the sets:
\[d(A,B)=\displaystyle\min_{1\le i\le N_A,1\le j \le N_B}\,d(a_i,b_j).\]

{\it Nearest-neighbor clustering} is a process that combines points from
$\mathbb{S}$ into consecutively larger subsets, called {\it clusters}, by connecting at each step
the two nearest clusters; this process can be described by the {\it nearest-neighbor
spanning tree} $\mathbb{T}$.
Specifically, consider $N$ points $c^0_i\in\mathbb{S}$, $i=1,\dots,N$ with pairwise
distances $d^0_{ij}\equiv d(c^0_i,c^0_j)$.
These points, considered as clusters of unit mass ($m_i=1$), form $N$ leaves of the
time-oriented tree $\mathbb{T}$.
The first internal tree node is formed at the time
$t_1=\min_{ij}\,d^0_{ij}$ by merging two closest points $c^0_{i^*}$ and $c^0_{j^*}$ with
$(i^*,j^*)={\rm argmin}_{ij}\,d^0_{ij}$, where
${\rm argmin}_{ij}\,f(i,j)$ is defined as a pair $(i^*,j^*)$ such that
$f(i^*,j^*)=\min_{ij}\,f(i,j)$.
This merging creates a new cluster of two points, with a mass of $m_i+m_j=2.$
Hence, at time $t_1$, there exist $N-1$ clusters:  $N-2$ clusters with unit mass and
one cluster of mass $m=2$.

We can now reindex the clusters so as to work with clusters $c^1_i$, $i=1,\dots,N-1$;
their total mass is $\sum_{i=1}^{N-1}\,m_i=N$ and pairwise distances are
$d^1_{ij}\equiv d(c^1_i,c^1_j)$.
The second internal node of tree $\mathbb{T}$ is formed at time $t_2=\min_{ij}\,d^1_{ij}>t_1$
by merging the two closest clusters from the set $\{c^1_i\}_{i=1,\dots,N-1}$.
Thus, at time $t_2$ we have $N-2$ clusters $c^2_i$ such that their total mass is $N$
and pairwise distances are $d^2_{ij}\equiv d(c^2_i,c^2_j)$.
We continue in the same fashion, so the $k$-th internal cluster, for $1\le k \le N-2$,
is formed at time $t_k=\min_{ij}\,d^k_{ij}>t_{k-1}$, and at that time we have $(N-k)$
clusters $c^k_i$, $i=1,\dots,N-k$ with masses
$m_i$ such that $\sum_{i=1}^{N-k}\,m_i=N$.
Finally, at time $t_{N-1}$ we create a single cluster of mass $N$ that combines all
points $c^0_i$; this cluster forms the root of the tree $\mathbb{T}$.

Consider two nodes $a$ and $b$ from the nearest-neighbor tree and
let $t_a$ and $t_b$ be their time marks; recall that the tree is time-oriented
by the definition of the successive times $t_k=\min_{ij}\,d^k_{ij}>t_{k-1}$ at
which the cluster mergers occur.
The {\it ancestors} of a node are its parent, the parent of that parent, and so on,
all the way to the root.
Clearly, the time mark for an ancestor is larger than that of a descendant.
The {\it nearest common ancestor} $p$ of nodes $a$ and $b$ is their common ancestor
with the minimal time mark $t_p$.

The distance $u(a,b)$ along the the nearest-neighbor tree is defined as
the maximum of the values $u(a,p)\equiv\ t_p-t_a$ and $u(b,p)\equiv\ t_p-t_b$.
This distance satisfies two of the usual distance axioms, symmetry and strict positivity,
but the triangle inequality can be replaced by a more stringent one, namely
\[u(a,b)\le \max\,[u(a,c),\,u(c,b)],\]
which holds for any three nodes $a,b$ and $c$.
Such a distance function is called an {\it ultrametric} [{\it Rammal et al.}, 1986; {\it Schikhof}, 2007].
Ultrametric spaces have many peculiar properties;
for instance, one can rename {\it any} triplet $a,b,c$ of nodes in such a way that
\[u(a,c)=u(b,c).\] These unusual properties give ultrametric spaces considerable
flexibility in applications, and point sets connected via nearest-neighbor
clustering are a representative example of such spaces.

In the billiard example of Section ~\ref{exs}, the space $\mathbb{S}$
is the set of $N$ billiard balls and the ultrametric distance function $u(a,b)$
equals the time before the first collision of the balls $a$ and $b$.
Naturally, their distance depends on the initial positions and velocities of the
two balls $a$ and $b$, but it is affected by the global billiard dynamics:
our two balls may be set to collide at a given time $t^*$ in the absence of other balls,
but may be hit by some other ball at time $t<t^*$, thus postponing the collision
of $a$ with $b$.

In our river transport problem, the space $\mathbb{S}$ is the set of all river sources.
The ultrametric distance $u(a,b)$ between two sources is defined as the time necessary
for the corresponding fluxes injected into these two sources to meet
down the river path.
If the static river geometry is described by the tree $\mathbb{T}_{\rm S}$ --- and
we assume, as previously stated, that fluxes move with unit speed downstream ---
the traditional distance $d(a,b)$ between two sources equals the maximal length
along the tree to their nearest common parent in $\mathbb{T}_{\rm S}$.
The nearest-neighbor spanning tree of hierarchical-aggregation theory becomes
what we called so far, in the context of river transport, the dynamic tree $\mathbb{T}_{\rm D}$.
As previously stated, this dynamic tree differs, in general, from the static tree
$\mathbb{T}_{\rm S}$ and depends not only on the topology of the latter, but also
on the actual length of the links.
If the velocities vary in time or space, then the spanning tree $\mathbb{T}_{\rm D}$
will depend on the specific dynamics of the processes operating on the static tree.

To better understand transport on river networks, we will elucidate in the next section
the connection between the statistical properties of $\mathbb{T}_{\rm S}$
and those of $\mathbb{T}_{\rm D}$.

\section{Analysis of drainage networks}
\label{rivers}

%\subsection{Research questions}
In this section we quantify similarities and differences
between the branching topology of static and dynamic trees, both at the stream and
hillslope network scale.
%Specifically, we examine the presence of self-similarity and the presence or absence
%of universality across three river basins of different physiography.
%We then examine the aggregation or cluster dynamics of the dynamic trees and interpret
%the critical time of cluster formation in relevance to transport or maximum connectivity
%along that river network

\subsection{Data description}
\label{data}
We have analyzed three river basins:
Upper Noyo (Mendocino County, California, USA), Tirso (Sardinia, Italy),
and Grigno (Trento, Italy).
Infomration about the physiographic and geologic characteristics of these basins
can be found in, respectively, {\it Sklar et al.} [2006], {\it Pinna et al.} [2004],
and {\it Guzzetti et al.} [2005].
The available DEMs were at a resolution of 10$\times$10 m$^2$ for the Noyo
basin, 30$\times$30 m$^2$ for the Grigno basin, and 100$\times$100 m$^2$
for the Tirso basin.
Since the focus of this study is not the extraction of the most
accurate river network from the available DEMs, we felt comfortable
adopting a simple criterion for channel initiation as 100 pixels
for all basins.
%this corresponds to a channel initiation area of $A_c=0.01$ km$^2$ for
%the Noyo, $A_c\approx 0.1$ km$^2$ for the Grigno and
%$A_c=1$ km$^2$ for the Tirso
Our main conclusions about the
comparison between the static and dynamic trees would not be affected
by changing the critical threshold areas within reasonable ranges.

\begin{figure}[p]
\centering\includegraphics[width=.5\textwidth]{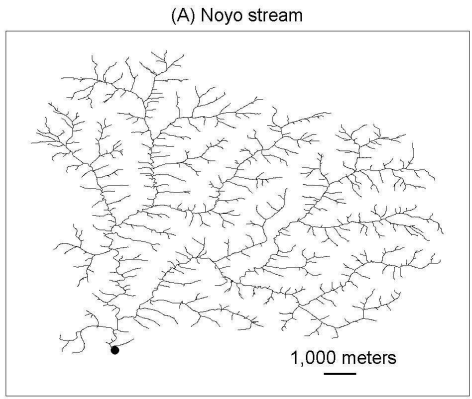}
\centering\includegraphics[width=.5\textwidth]{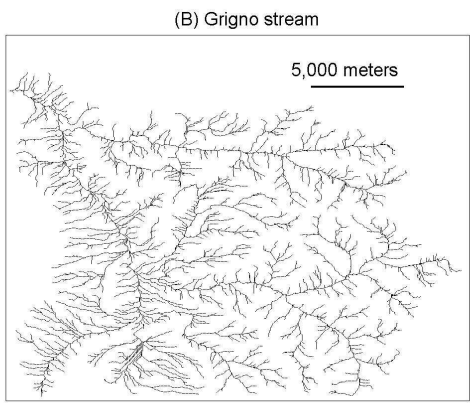}
\centering\includegraphics[width=.5\textwidth]{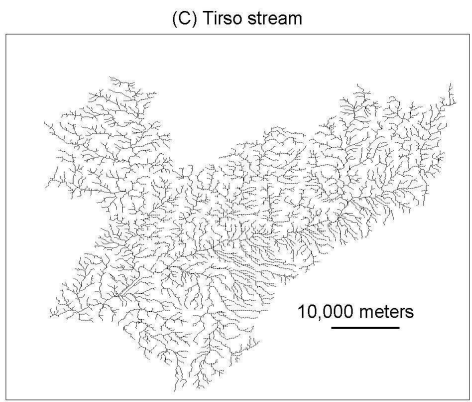}
\caption{Static trees for the stream networks of the three basins
analyzed in this study.
a) Noyo, Mendocino County, California, USA; the outlet is marked by a ball;
b) Grigno, Trento, Italy;
c) Tirso, Sardinia, Italy.
See Section ~\ref{data} for details of channel initiation.}
\label{fig_stream}
\end{figure}

The static trees we extracted from these DEMs for the three stream networks
are shown in Fig.~\ref{fig_stream}.
Using the procedure described earlier, we also extracted the static trees for
the hillslope networks, which drain every pixel of a basin, by using
a steepest gradient algorithm.
The corresonding dynamic stream and hillslope trees were then constructed
for each basin, assuming a constant unit speed of downstream propagation
for the fluxes.
Thus, we analyzed four different kinds of tree --- static stream, dynamic stream,
static hillslope, and dynamic hillslope -- for each basin.

%\subsection{Research questions}

%The questions we pose in this study and the short answers we obtain are:
%\begin{itemize}
%\item[(1)] Are the static stream and hillslope trees self-similar? [Yes, they are Tokunaga SSTs.]
%\item[(2)] Do the static stream trees differ from the static hillslope trees? [Yes.]
%\item[(3)] Are the dynamic stream and hillslope trees self-similar? [Yes, they are Tokunaga SSTs.]
%\item[(4)] Do the dynamic stream trees differ from the dynamic hillslope trees? [Yes.]
%\item[(5)] Do the dynamic trees differ from the corresponding static trees? [Yes.]
%\item[(6)] Are the topologies of four classes of trees universal
%or basin-dependent? [Largerly universal, with some basin-dependent variability.]
%\item[(7)] In a downstream transport, is there a critical time necessary for the
%flux to ``fill the stream network''? [Yes, the stream network is largerly filled
%after a critical time, similarly to the percolation process.]
%\end{itemize}

\subsection{Self-similar properties}

\begin{figure}[p]
\centering\includegraphics[width=.45\textwidth]{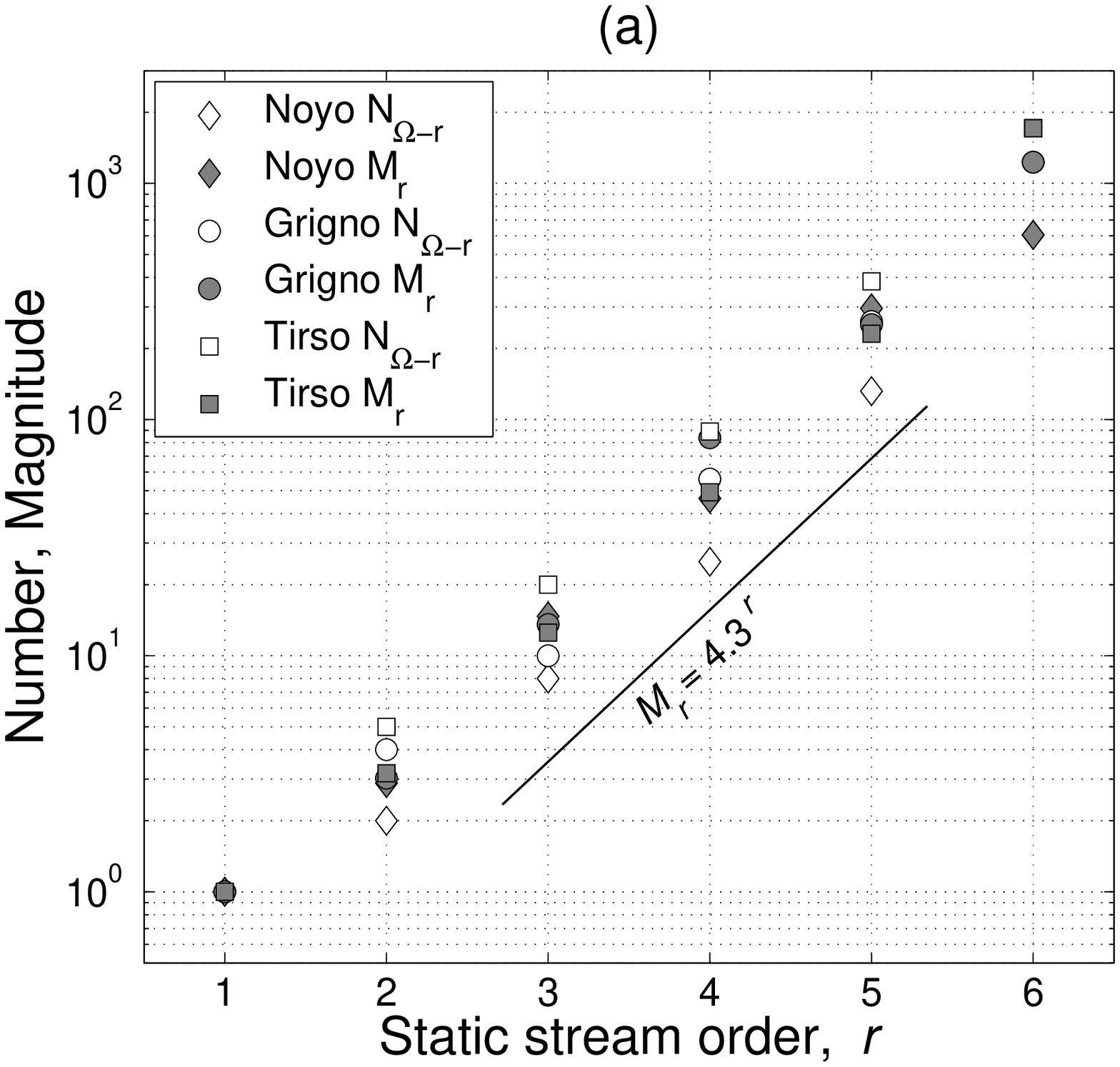}
\centering\includegraphics[width=.45\textwidth]{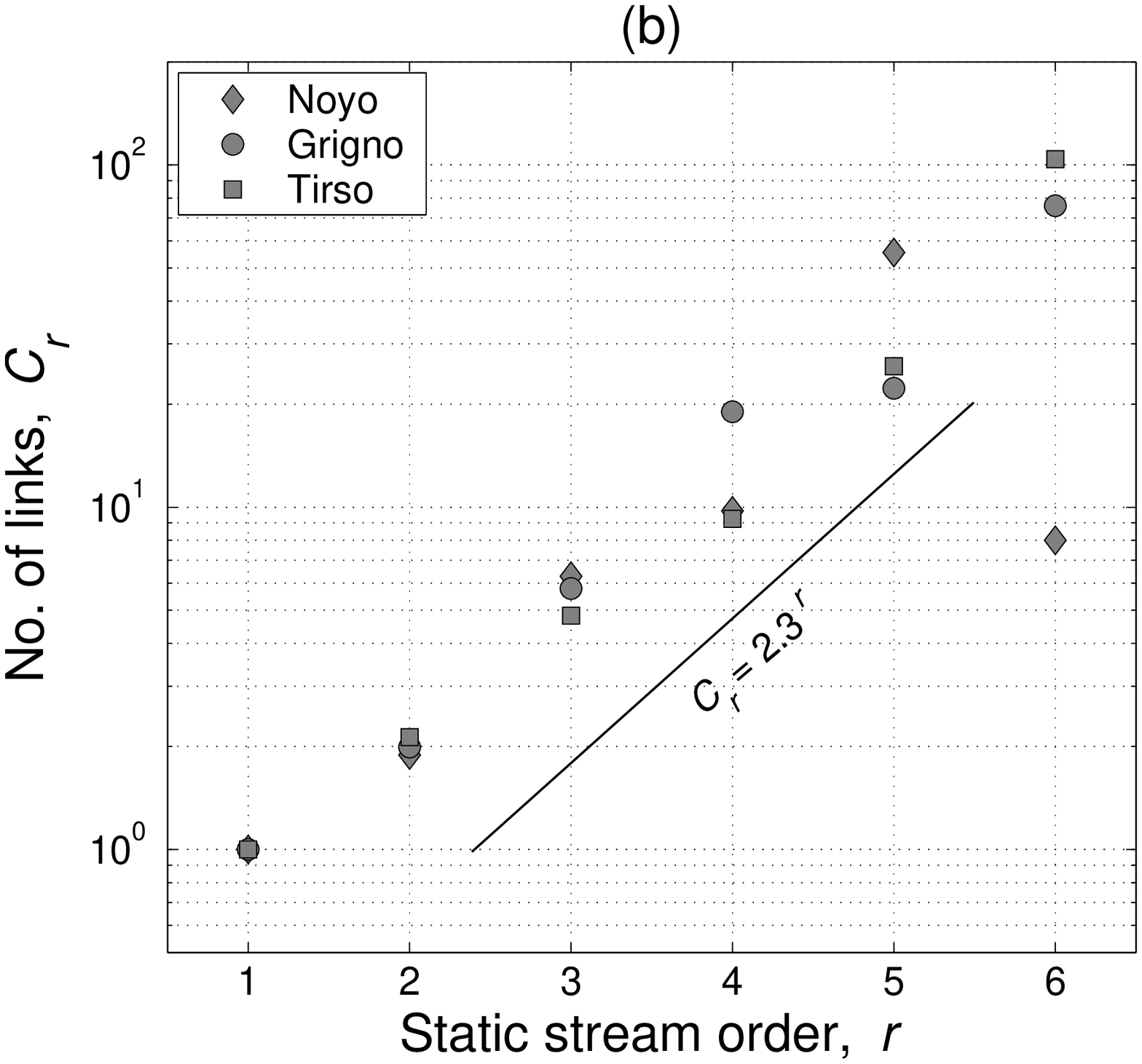}
\centering\includegraphics[width=.45\textwidth]{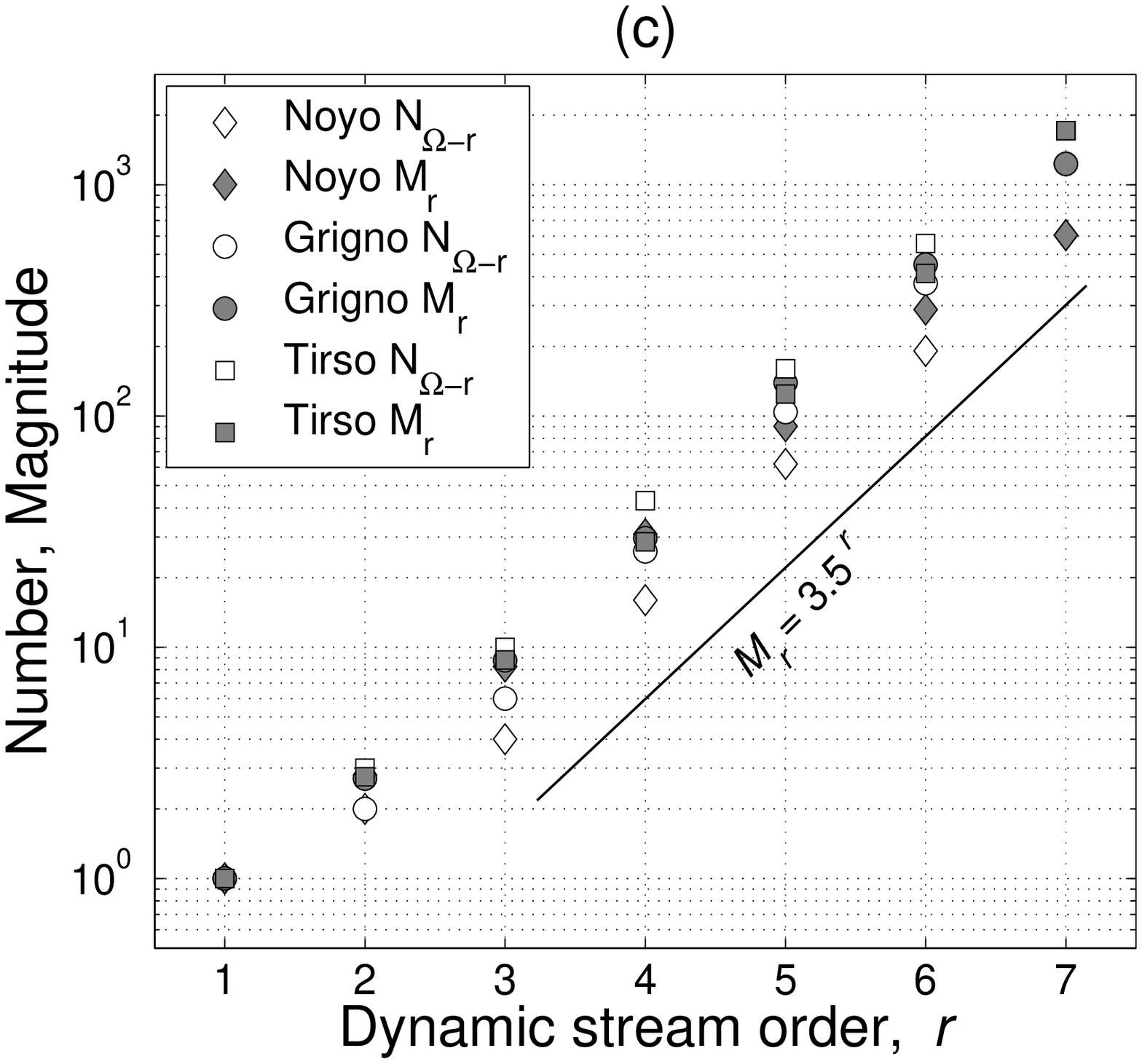}
\centering\includegraphics[width=.45\textwidth]{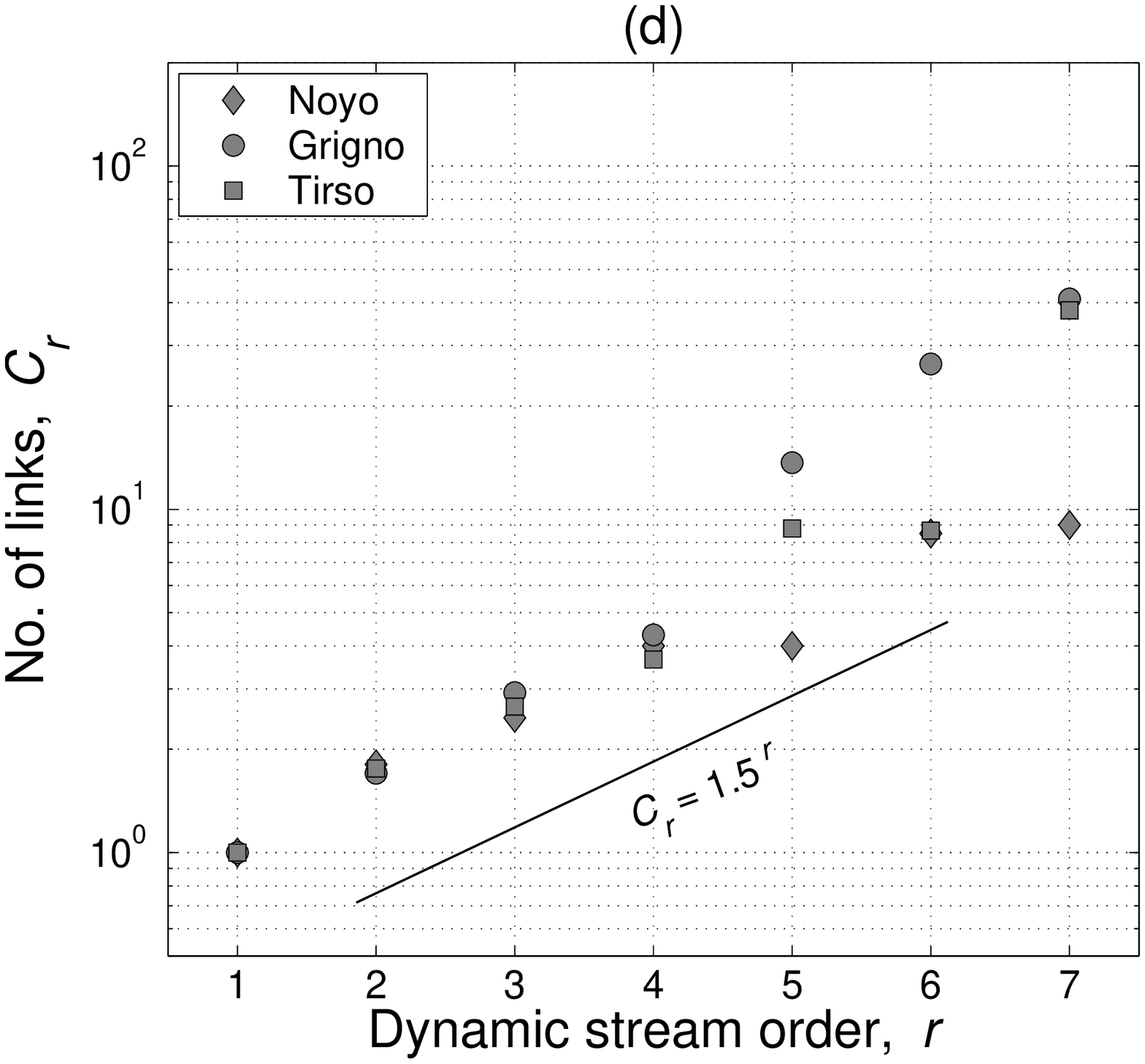}
\caption{Branching statistics for the stream trees
of Noyo, Grigno, and Tirso basins.
Number $N_r$ and average magnitude $M_r$ for static (panel a)
and dynamic (panel c) trees and
average number $C_r$ of links within a branch for static
(panel b) and dynamic (panel d) trees.}
\label{fig_NM_S}
\end{figure}

\begin{figure}[p]
\centering\includegraphics[width=.45\textwidth]{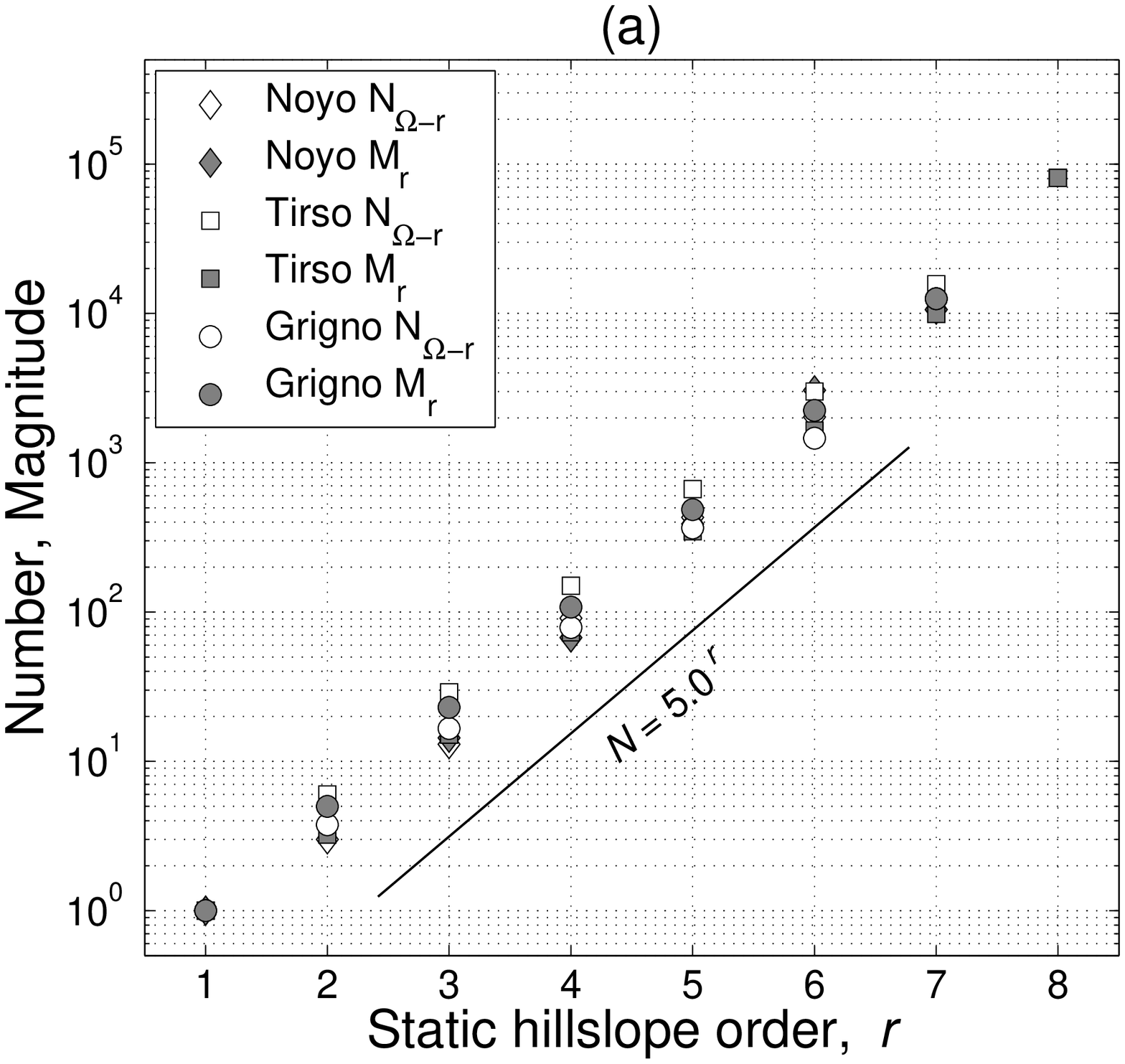}
\centering\includegraphics[width=.45\textwidth]{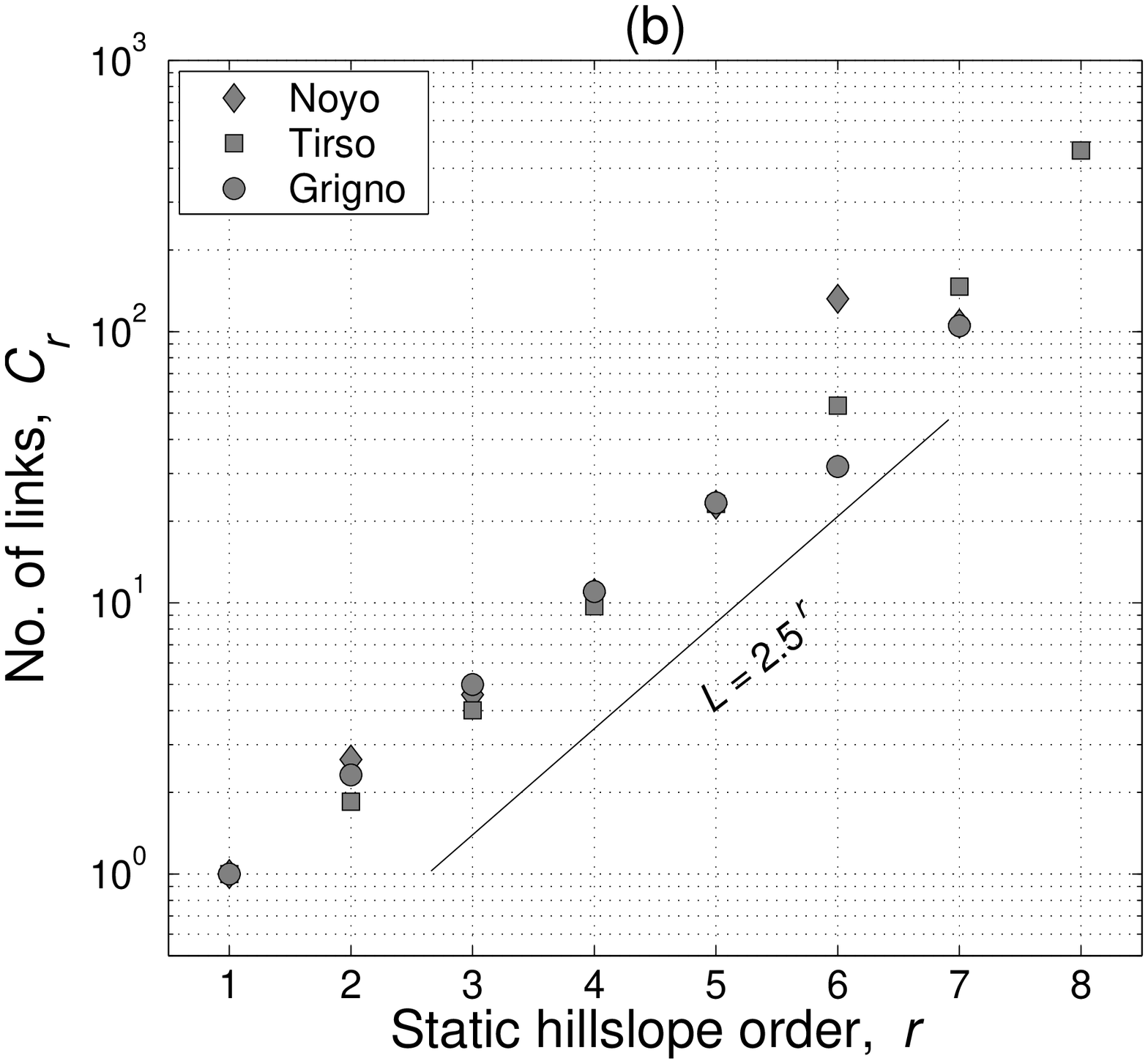}
\centering\includegraphics[width=.45\textwidth]{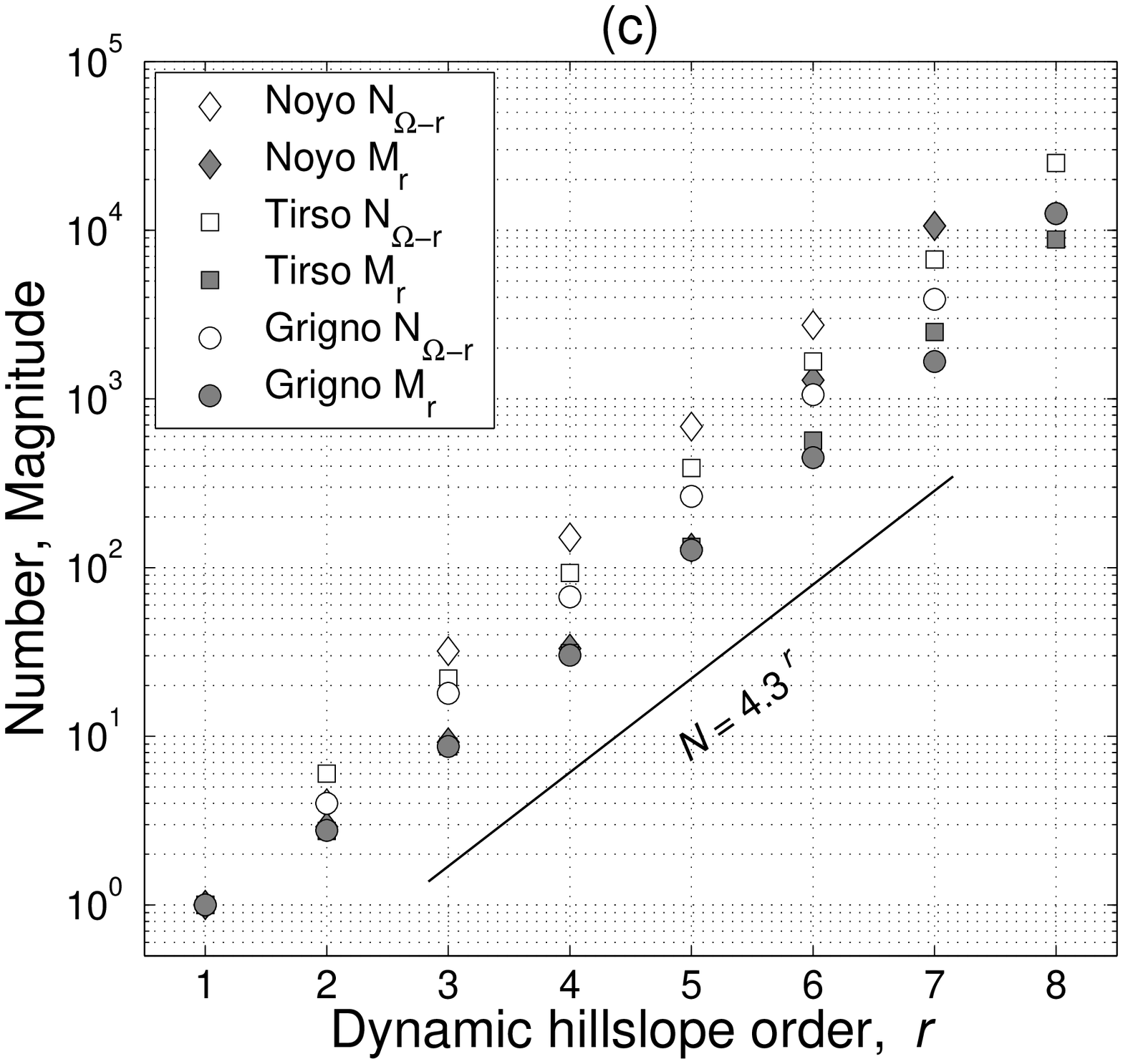}
\centering\includegraphics[width=.45\textwidth]{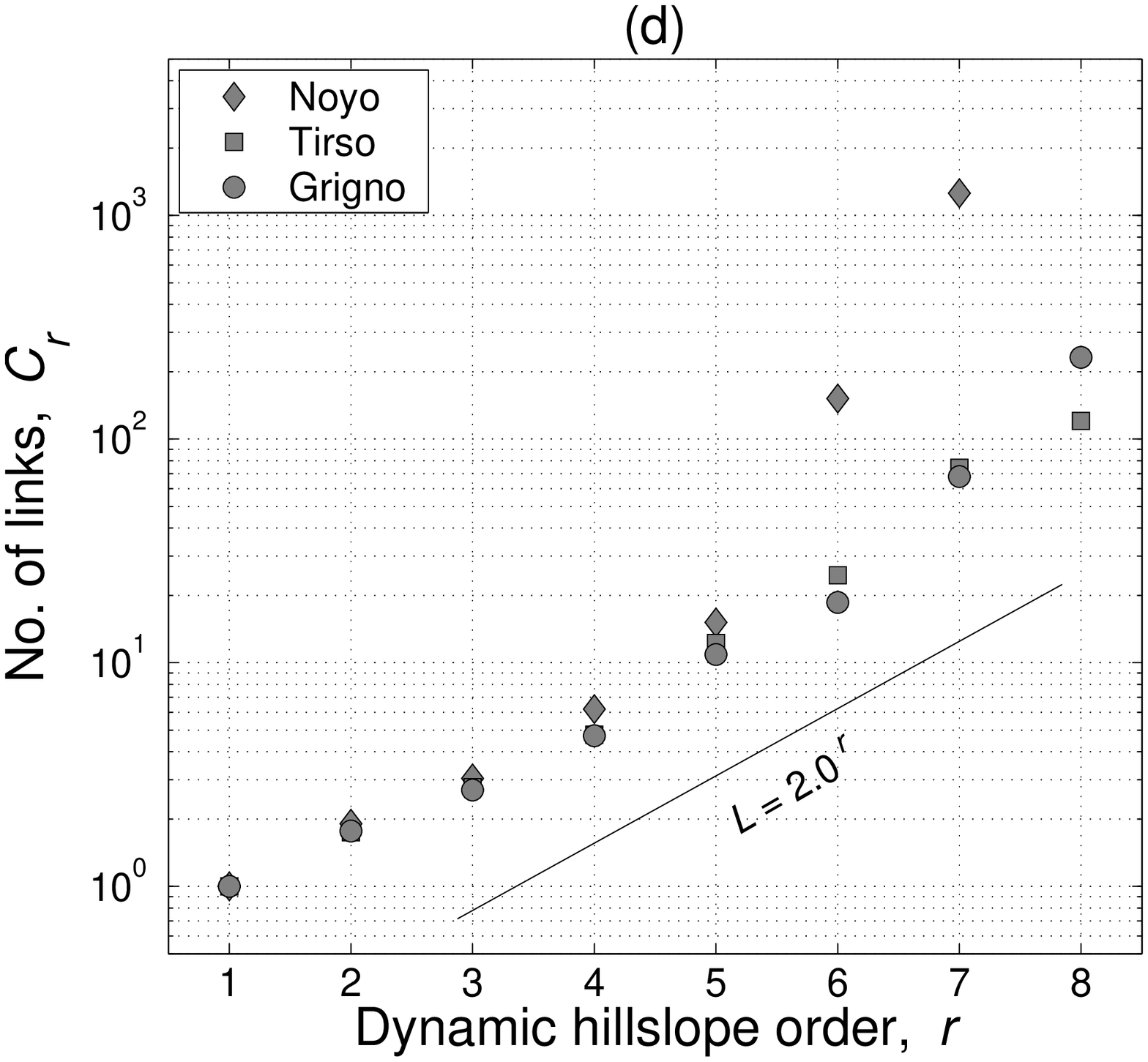}
\caption{Branching statistics for the hillslope trees
of Noyo, Grigno, and Tirso basins.
Number $N_r$ and average magnitude $M_r$ for static (panel a)
and dynamic (panel c) trees and
average number $C_r$ of links within a branch for static
(panel b) and dynamic (panel d) trees.}
\label{fig_NM_H}
\end{figure}

Figures~\ref{fig_NM_S} and \ref{fig_NM_H} show the distributions of the number $N_r$,
average magnitude $M_r$, and the average number $C_r$ of links for branches of
order $r$.
The results in Fig.~\ref{fig_NM_S} refer to the stream trees;
the results in Fig.~\ref{fig_NM_H} to the  hillslope trees.

Despite the small-sample fluctuations, the figures demonstrate a large degree
of consistency among the branching indices for the trees from different classes.
All considered branching statistics are closely approximated by the Horton laws.
Moreover, the results suggest that the relationship \eqref{ratios}
holds in all the considered cases.
Furthermore, we observe that the values of the stream ratios for static trees are
higher than the corresponding values for dynamic trees; and the values of the
stream ratios for stream trees are smaller than the corresponding values for
hillslope trees.

The only indices that considerably deviate from the Horton laws at higher orders
are $C_r$ (average number of nodes within an order-$r$ branch) for the Noyo basin
and this warrants special investigation in the future.
Apart from this discrepancy, overall we conclude that the four classes of trees,
dynamic vs. static and stream vs. hillslope, can be closely approximated by the Tokunaga SSTs.

\subsection{Phase transition in hierarchical dynamics}
\label{PT}

\begin{figure}[p] %[p] [t]
\centering\includegraphics[width=.45\textwidth]{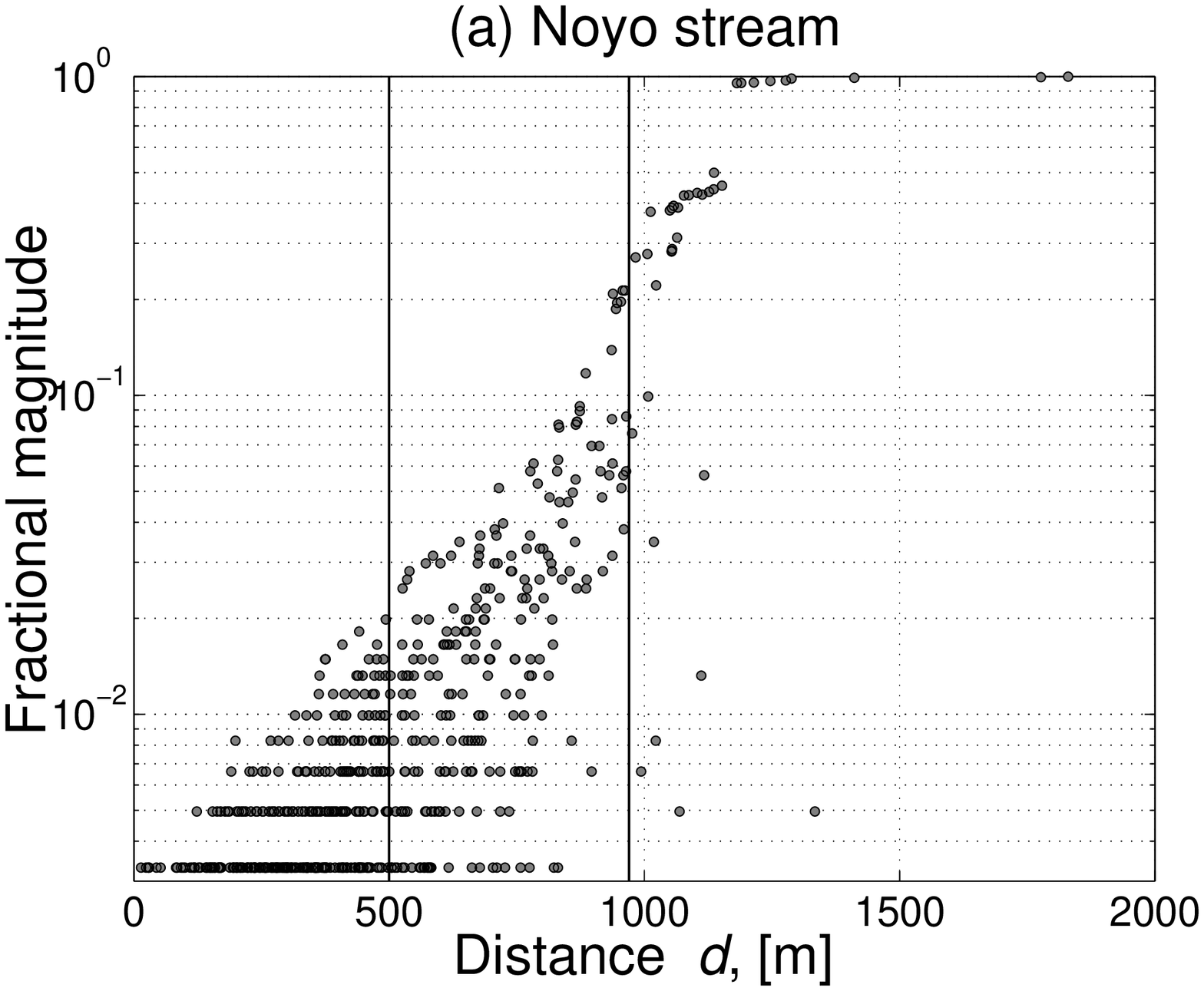}
\centering\includegraphics[width=.45\textwidth]{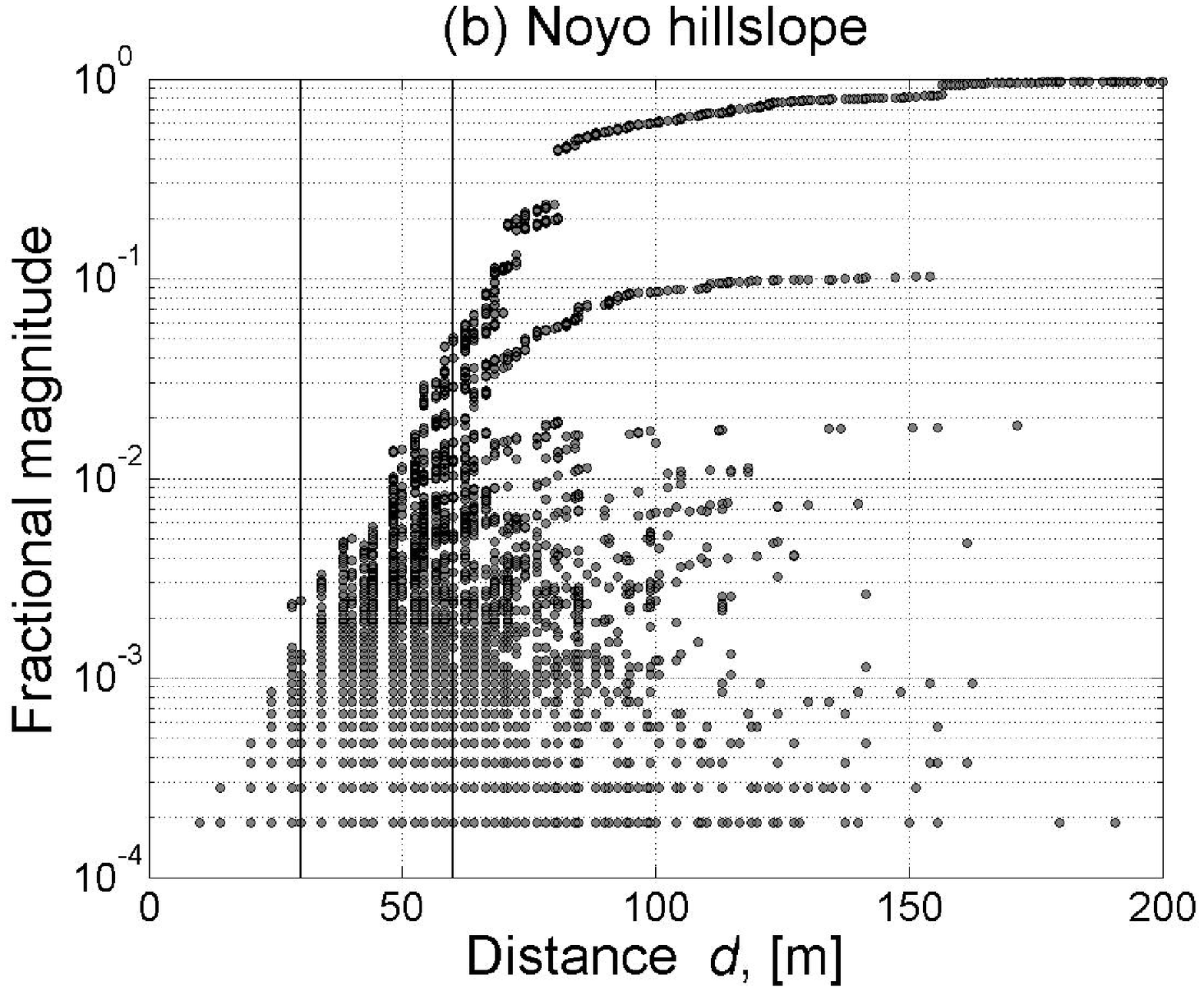}
\centering\includegraphics[width=.45\textwidth]{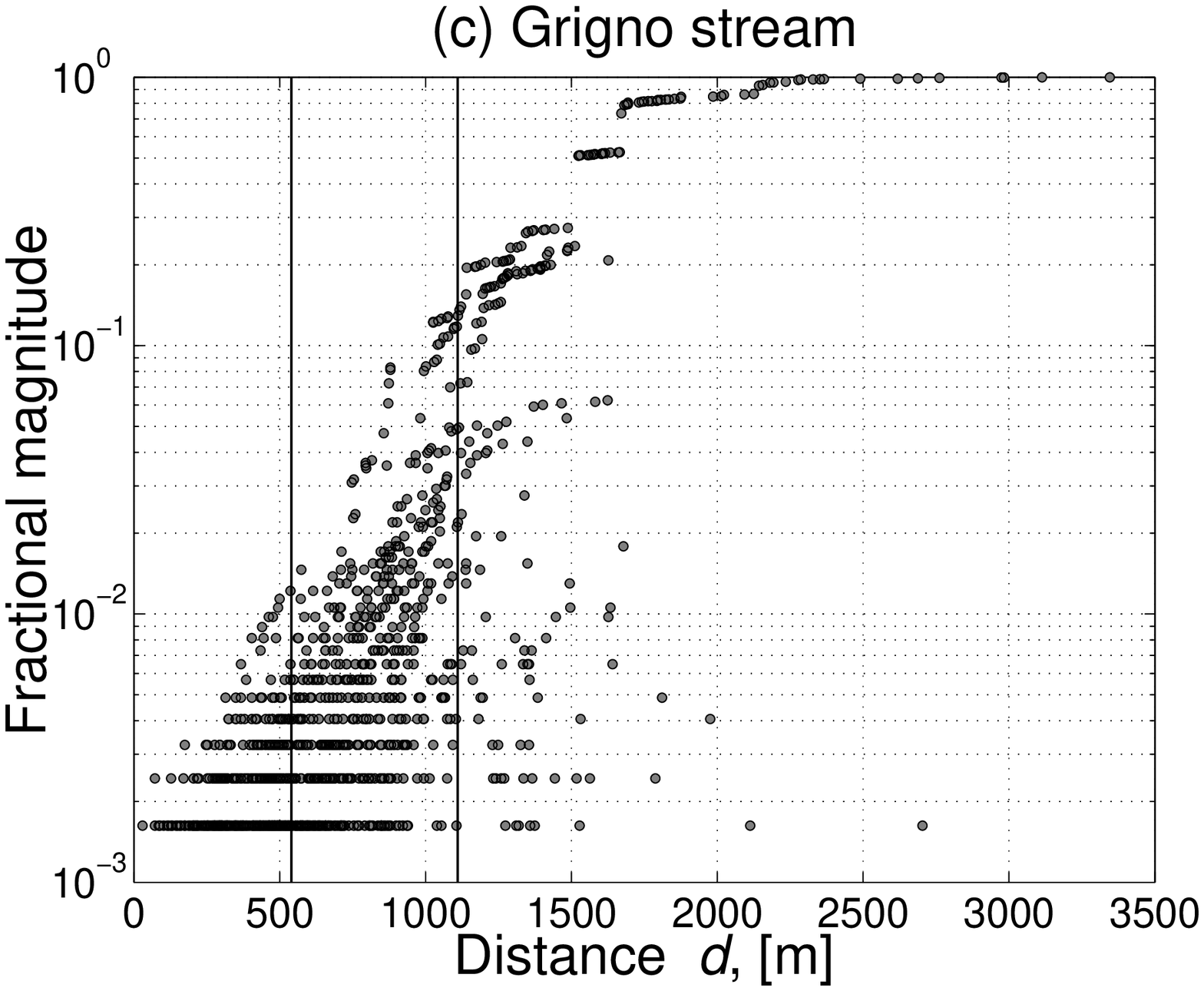}
\centering\includegraphics[width=.45\textwidth]{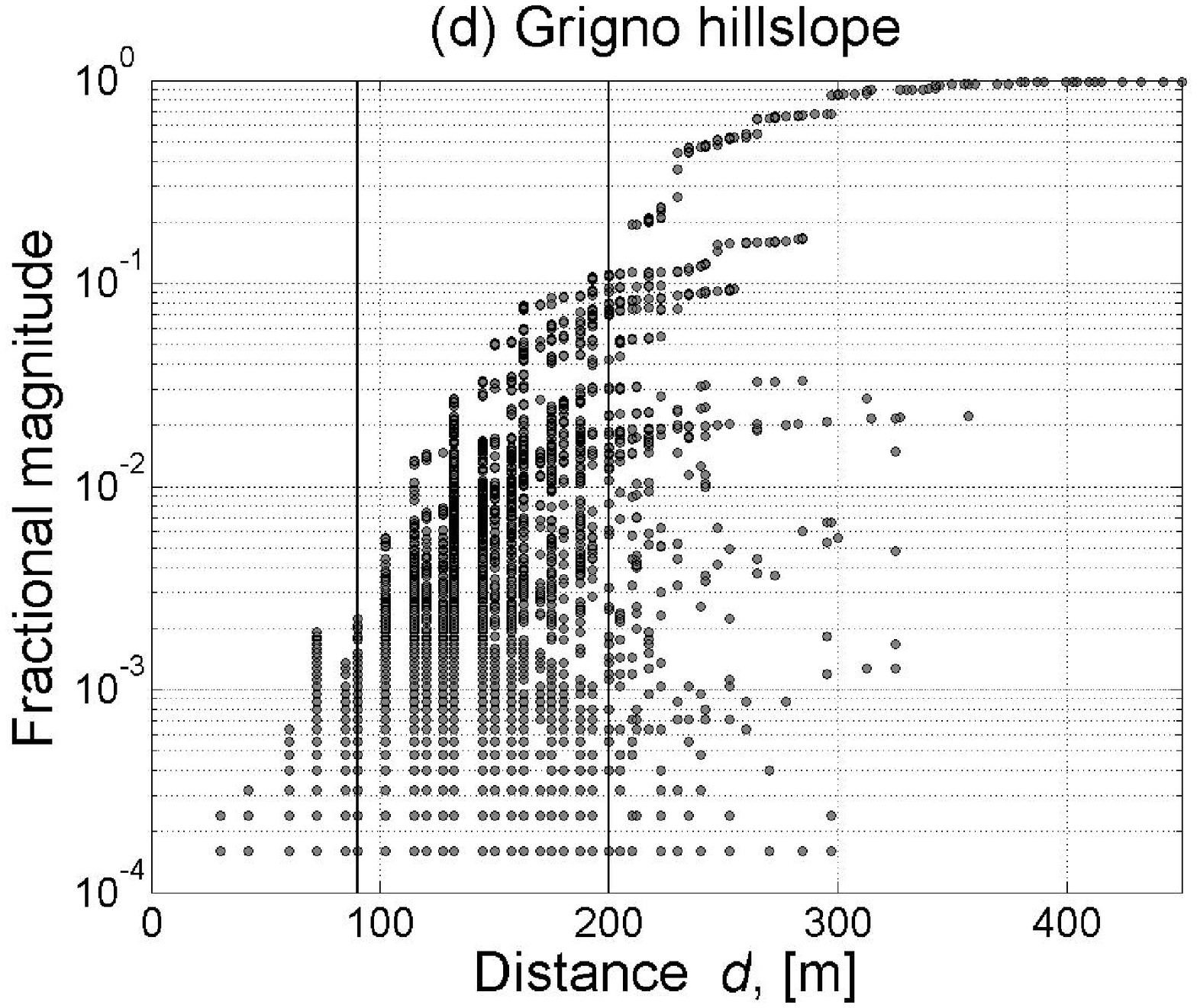}
\centering\includegraphics[width=.45\textwidth]{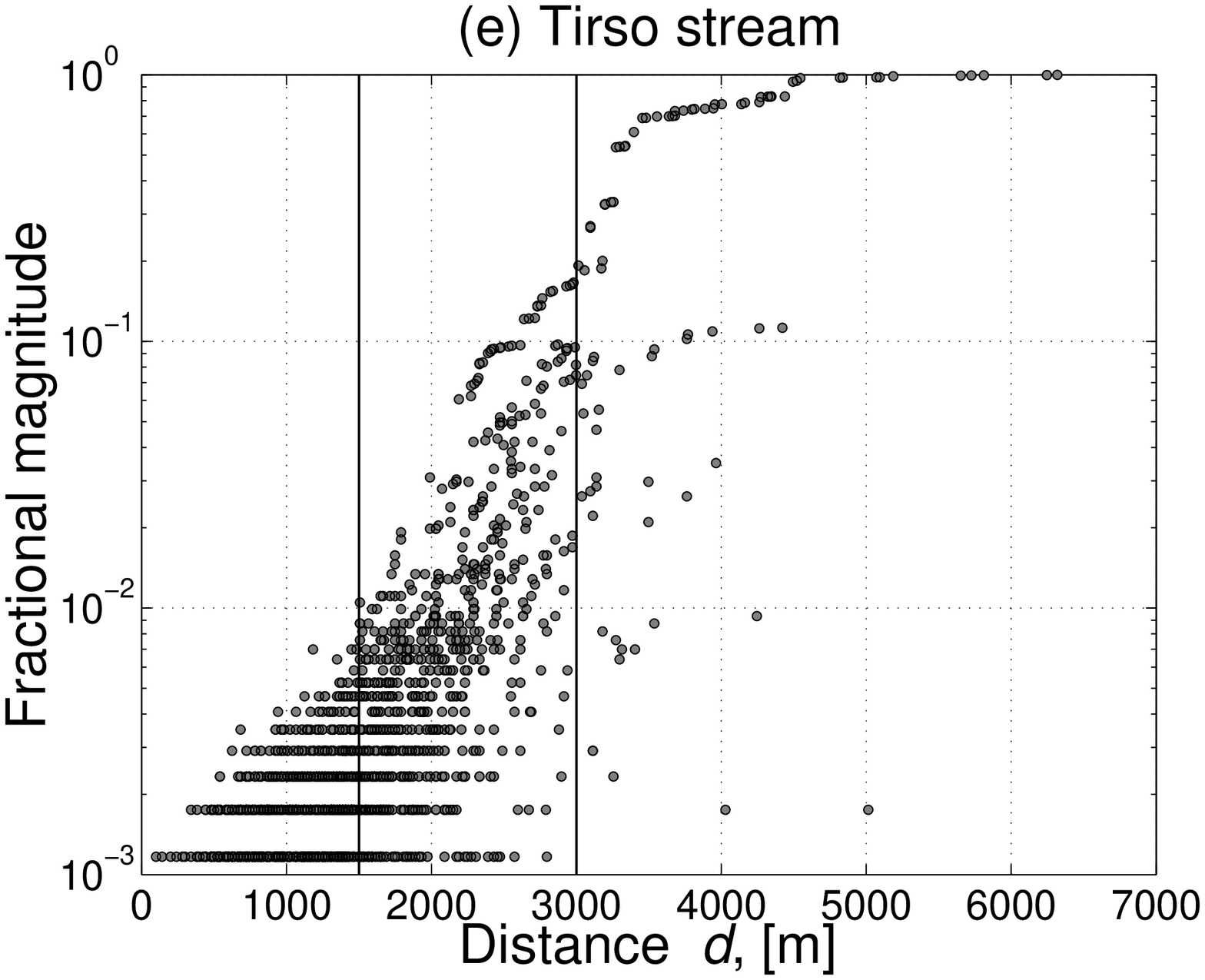}
\centering\includegraphics[width=.48\textwidth]{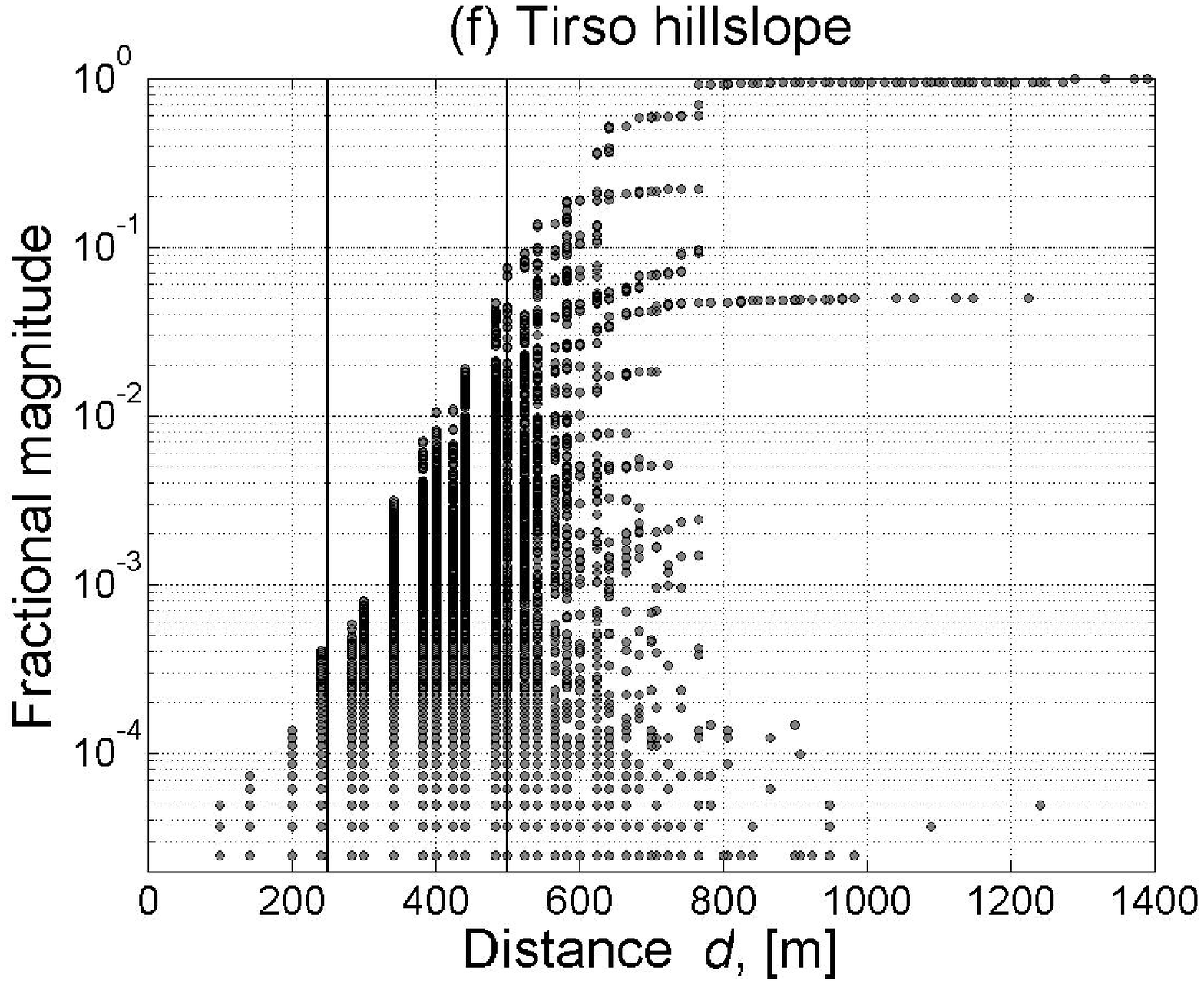}
\caption{Fractional branch magnitudes $m_i/N$ as a function of the distance
$d_i$ traveled by the dye at the branch creation instant.
Notice that the distance $d_i$ can be interpreted as the time $t_i$ necessary
to create the branch.
a) Noyo stream;
b) Noyo hillslope;
c) Grigno stream;
d) Grigno hillslope;
e) Tirso stream;
f) Tirso hillslope.}
\label{fig_PT}
\end{figure}

Here we ask the question as to whether the river network connectivity (in terms of
elements of the network participating in transport) exibits a {\it phase transition}
akin to those found in other systems.
Figure~\ref{fig_PT} shows the fractional magnitudes $m_i/N$ of the branches in the
dynamic trees (stream and hillslope trees of the three river basins) as a function of the distance $d$ traveled by the dye.
Recall that this distance can also be interpreted as the time $t$ when the
node was created by merging of upstream branches.
Altogether we consider six cases; in all of them one observes the following
scenario.
We start at distance $d=0$ (or time $t=0$) with $N$ branches (clusters) of unit
magnitude corresponding to the most outer nodes of the transport tree.
As distance increases (time evolves), the number of clusters decreases while their
magnitudes become larger and exhibit prominent variability.
In particular, at the small distances the maximal magnitude increases exponentially
with distance; this is reflected by an approximately linear form of an upper envelop
of the points in the figures (the envelop is not shown).
Furthermore, we notice that at the small distances (times) the magnitude distribution
is ``continuous'' in a sense that it does not have significant gaps.
However, at some critical time $t^*$ (translated here to distance $d^*$ for easier interpretation),
the distribution undergoes a serious qualitative
change: a prominent maximal cluster appears, such that its magnitude becomes significantly
larger than that of the second larger cluster.
Moreover, while the magnitude of the largest cluster keeps growing, the rest of
the distribution is fading off so after some time all clusters present at
$d=0$ merge with the largest cluster.
An interesting observation is that at the critical distance $d^*$ the magnitude of the
largest cluster is just about 10\% of the total magnitude $N$ of the system.
Notably, this number is universal for all the considered examples.

\begin{figure}[p] %[p] [t]
\centering\includegraphics[width=.35\textwidth]{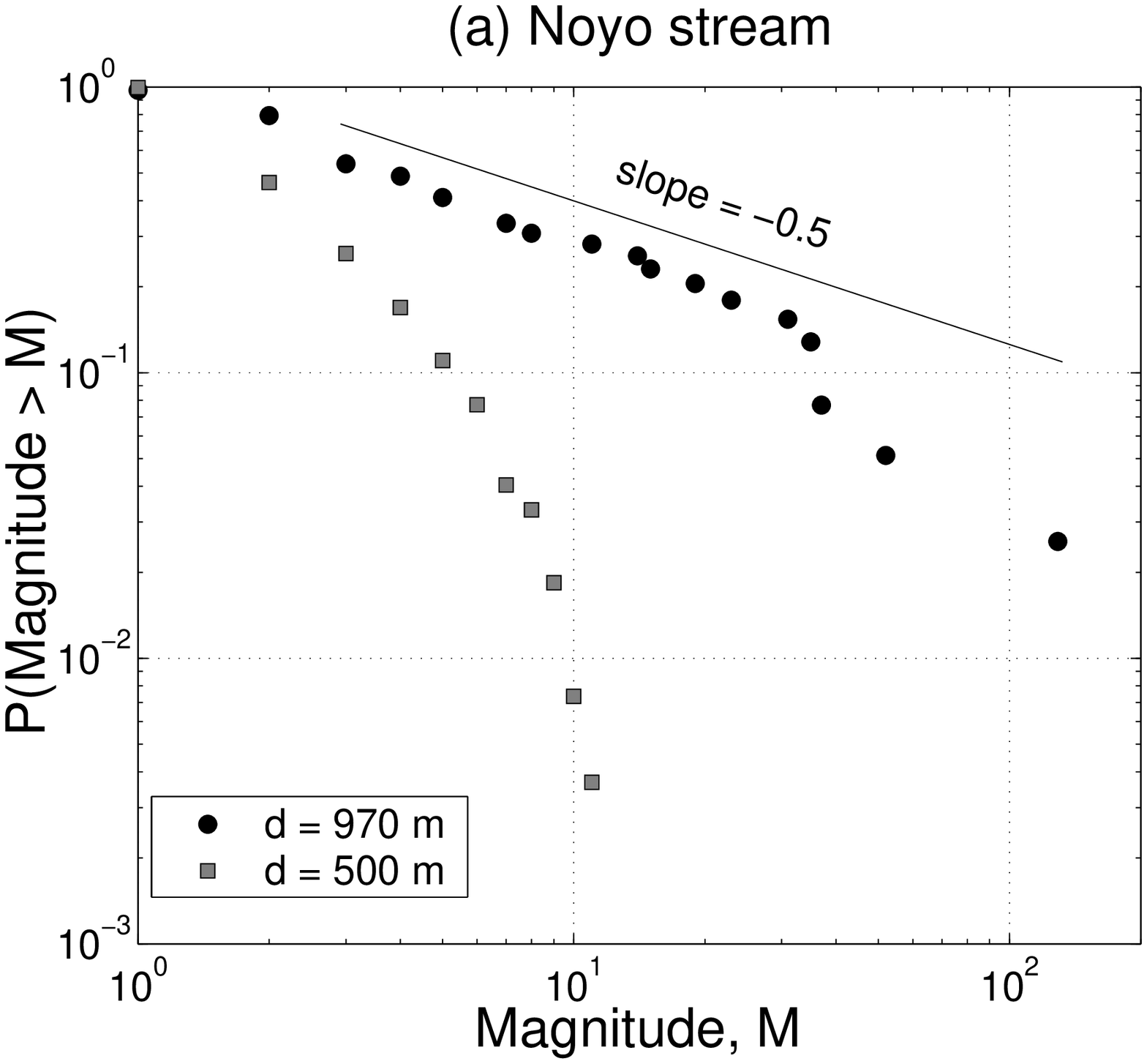}
\centering\includegraphics[width=.35\textwidth]{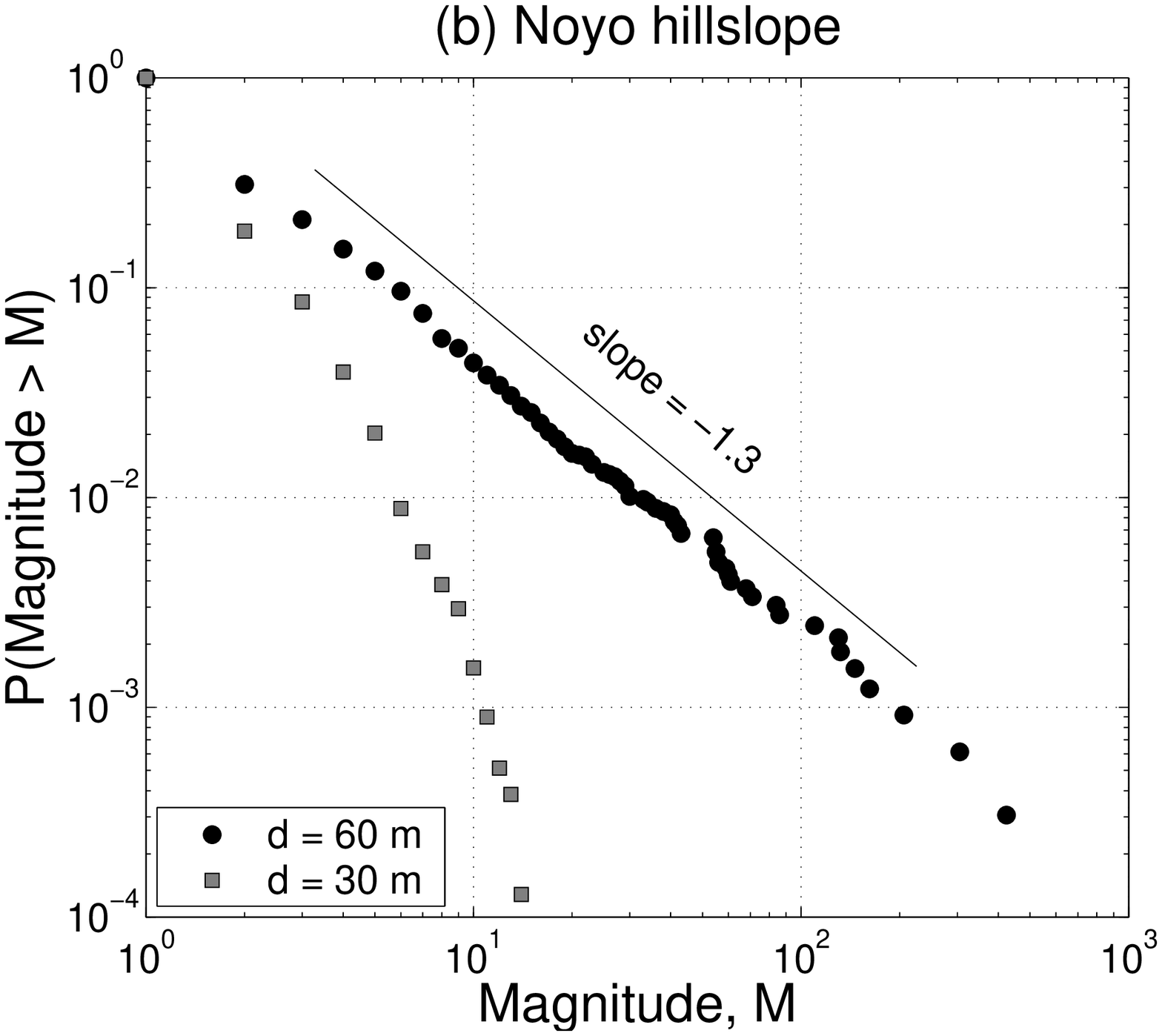}
\centering\includegraphics[width=.35\textwidth]{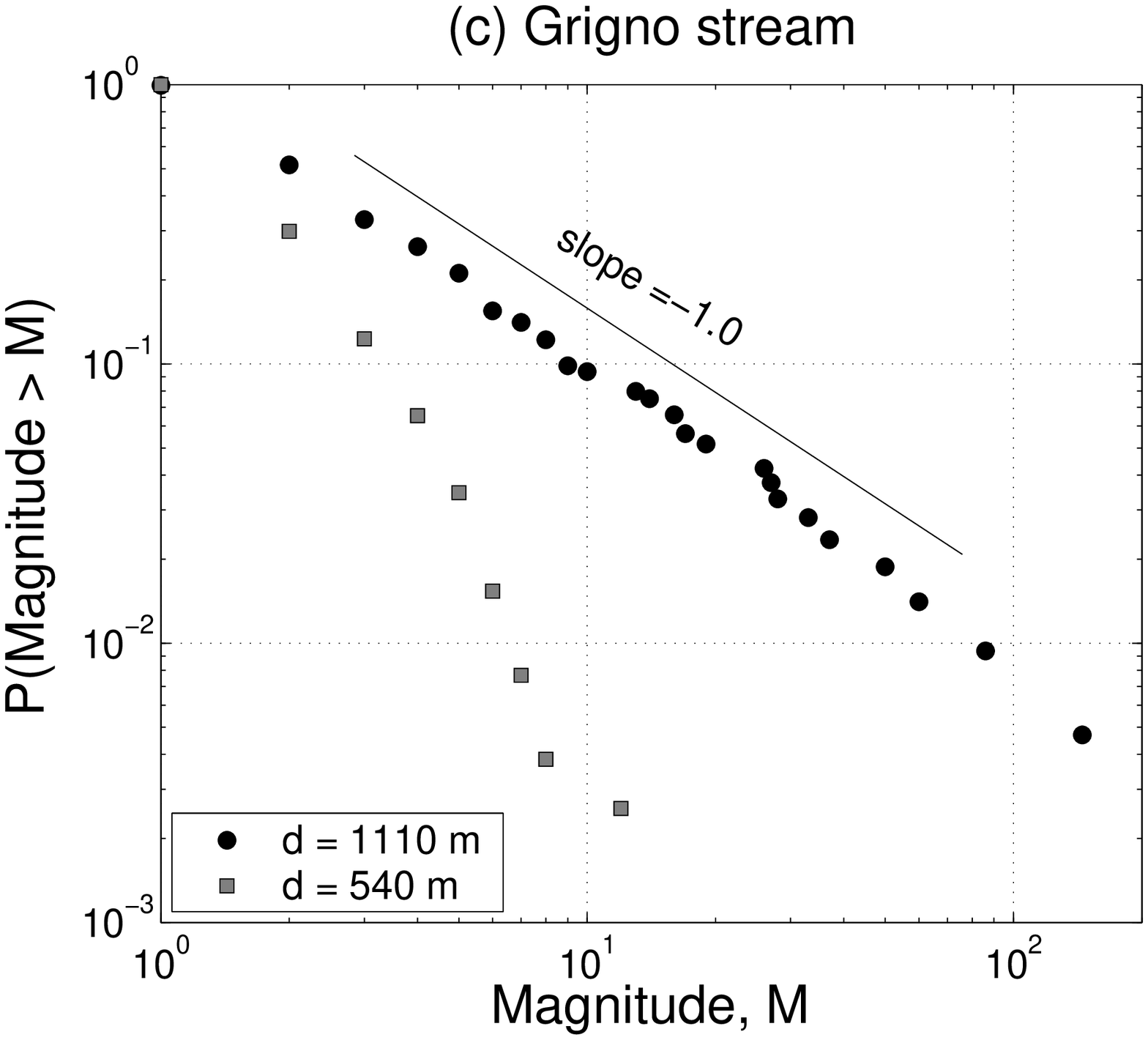}
\centering\includegraphics[width=.35\textwidth]{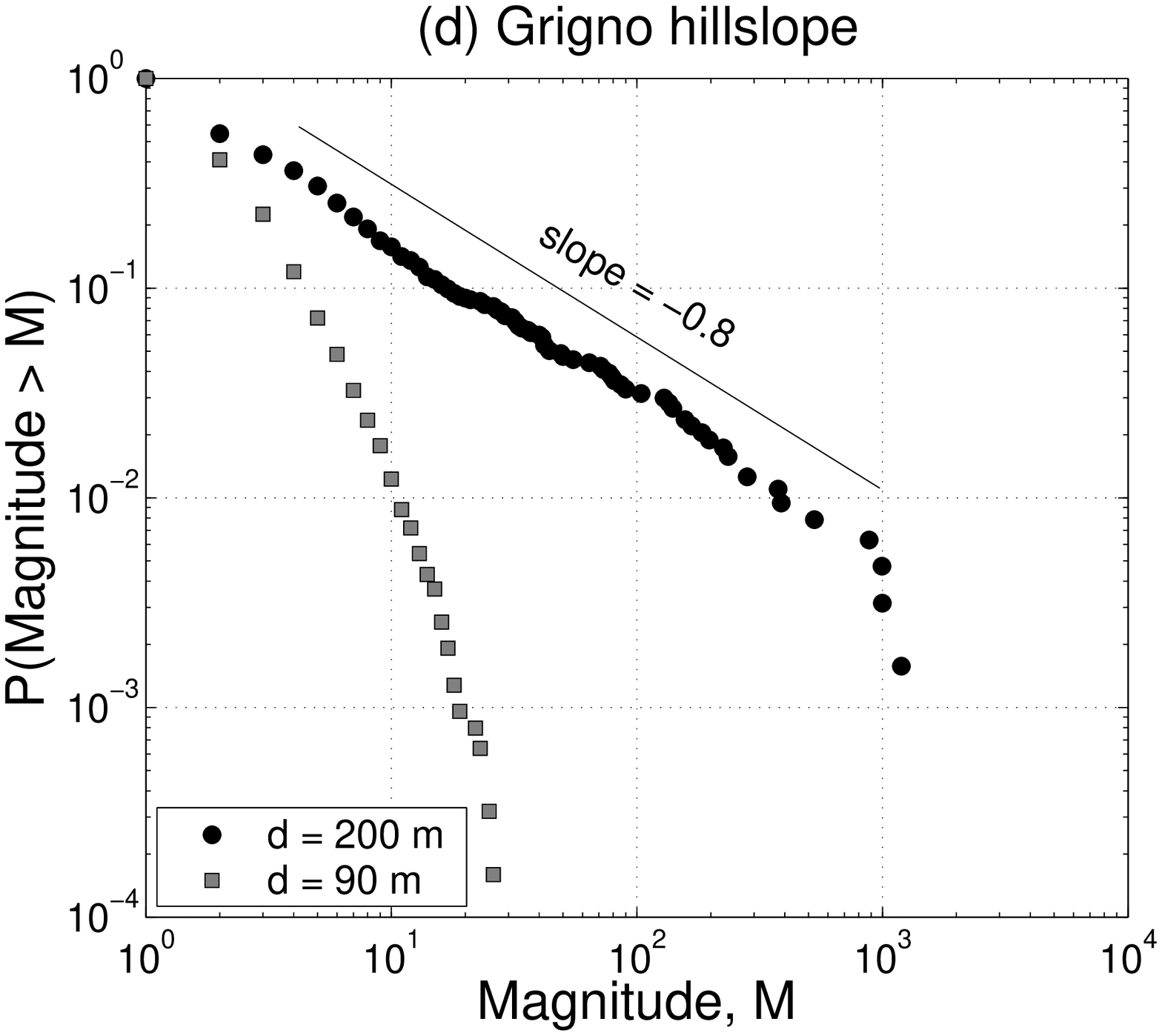}
\centering\includegraphics[width=.35\textwidth]{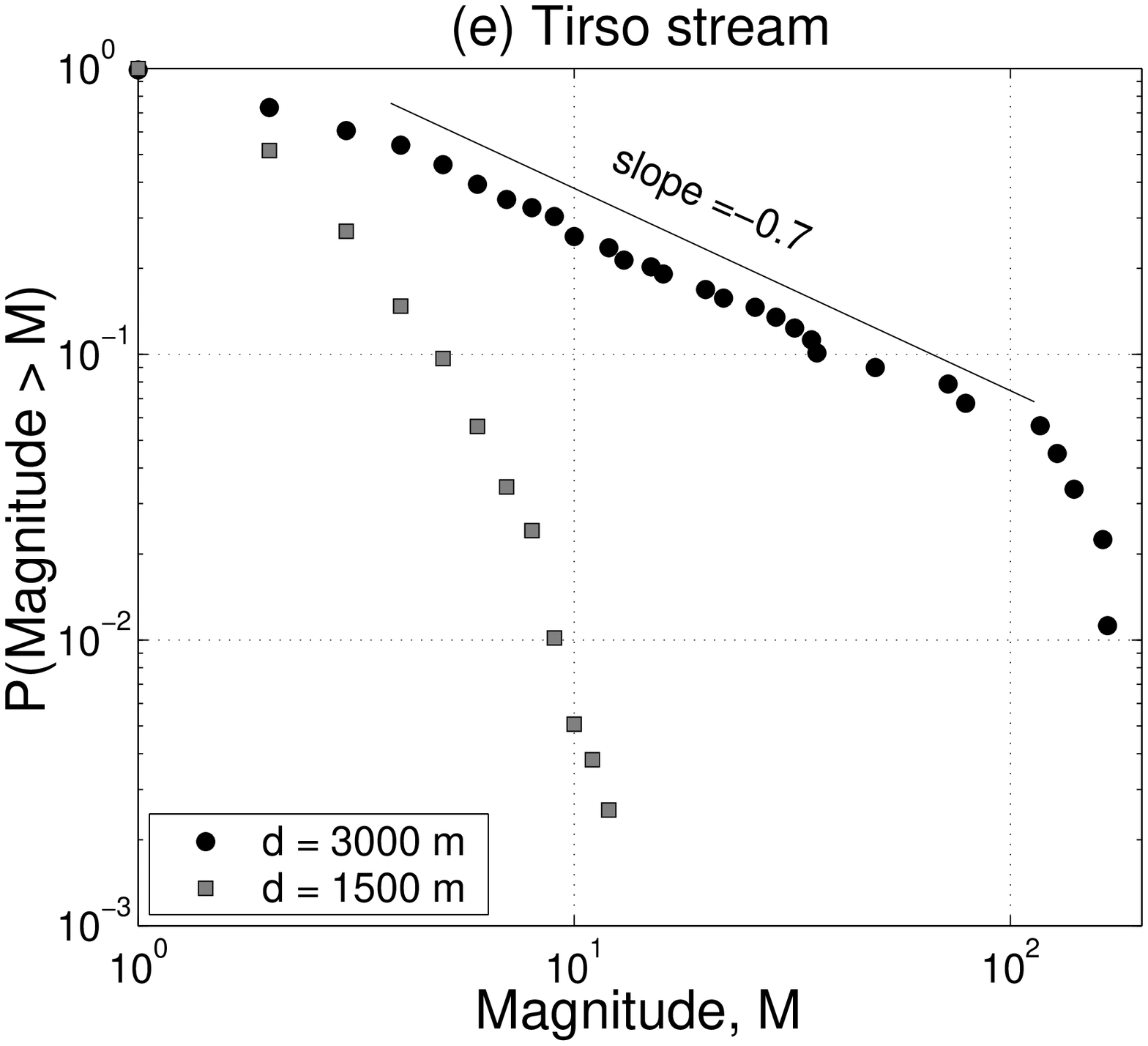}
\centering\includegraphics[width=.35\textwidth]{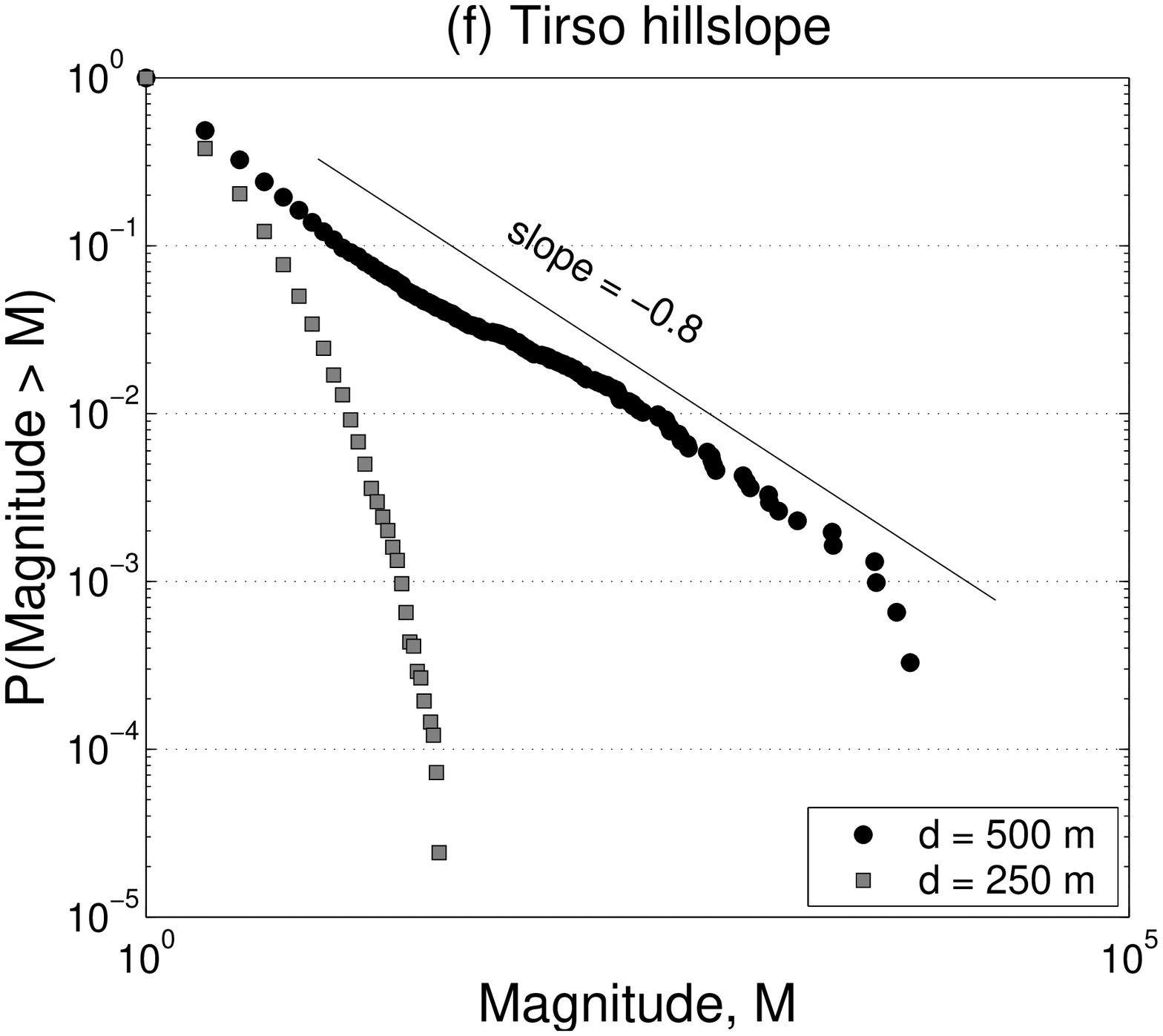}
\caption{Distribution of branch magnitudes $m_i$ at the critical distance $d^*$
(balls) and at an earlier time (squares) in dynamic trees for the three
basins.
a) Noyo stream;
b) Noyo hillslope;
c) Grigno stream;
d) Grigno hillslope;
e) Tirso stream;
f) Tirso hillslope.
Each panel shows two distributions, the corresponding distances are depicted
by vertical lines in Fig.~\ref{fig_PT}.
Notice that the value of the critical distance $d^*$ can be interpreted as
the critical time $t^*$ necessary to create the critical cluster.
The downward deviations from the pure power laws are due to the finite-size
effect.}
\label{fig_GR}
\end{figure}

Figure~\ref{fig_GR} shows the magnitude distribution of the clusters that
existed when the dye traveled a given distance $d$.
The analysis is done for the critical distance $d^*$ and a smaller distance $d\approx d^*/2$;
they are both indicated by vertical lines in Fig.~\ref{fig_PT}.
In all six cases, we see that the magnitude distribution at the smaller distance (squares)
has an exponential tail, while at the critical distance (circles) it is a power law.
Recall that, in a log-log plot, power-law behavior shows up as
a straight line, while exponential behavior becomes a convex curve.
This change indicates that a phase transition occurs at the distance $d^*$.

\begin{figure}[p] %[p] [t]
\centering\includegraphics[width=.4\textwidth]{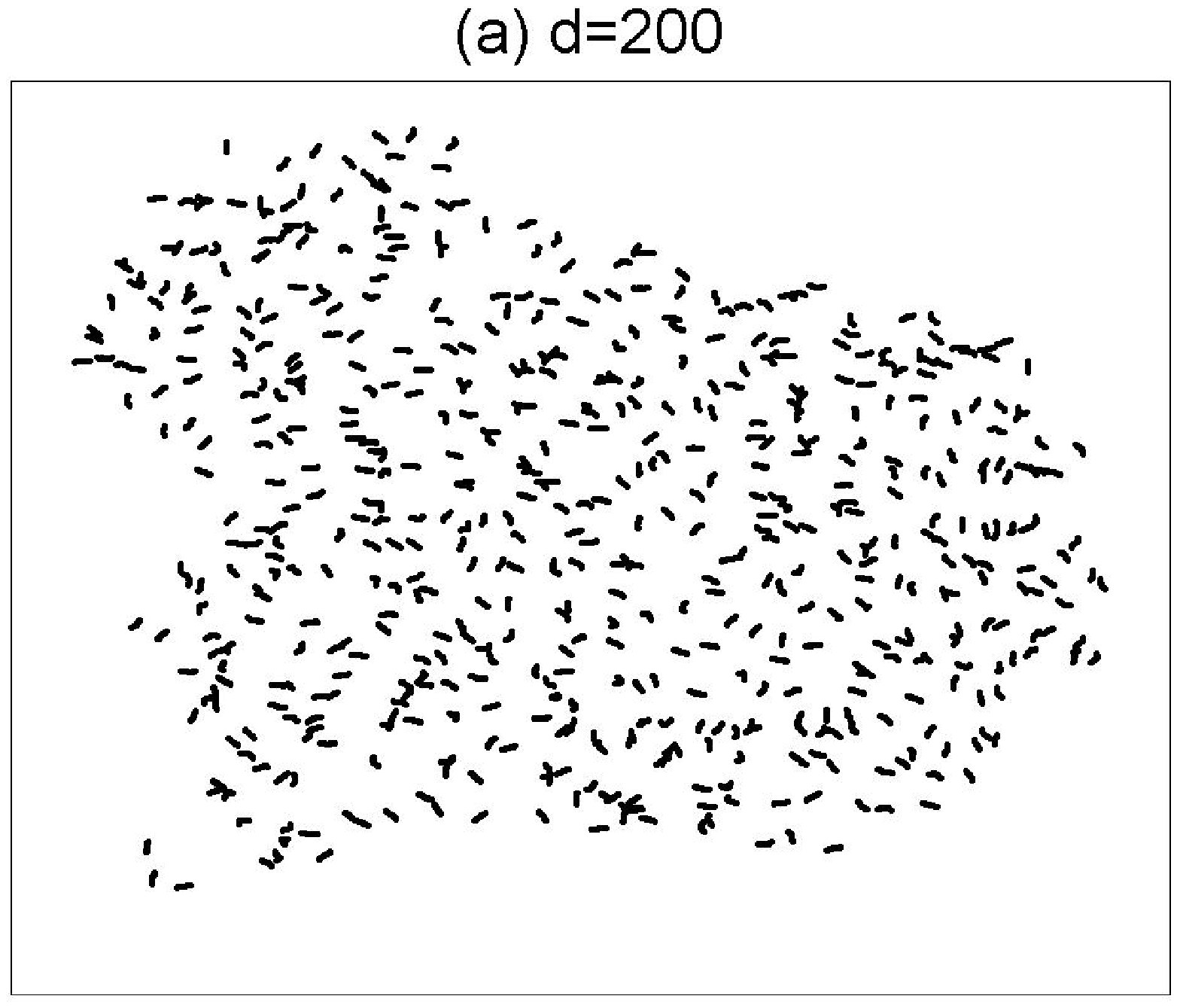}
\centering\includegraphics[width=.4\textwidth]{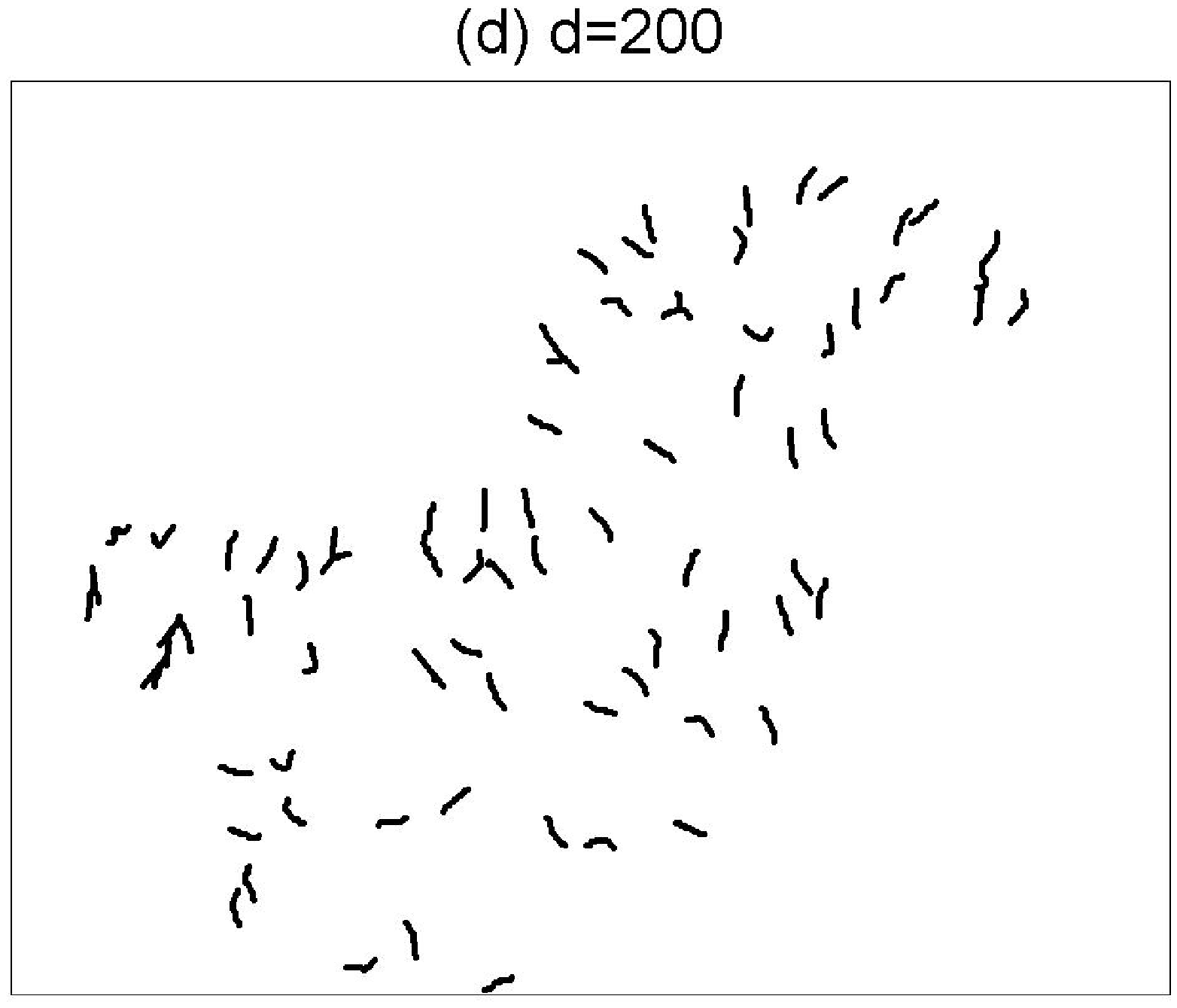}
\centering\includegraphics[width=.4\textwidth]{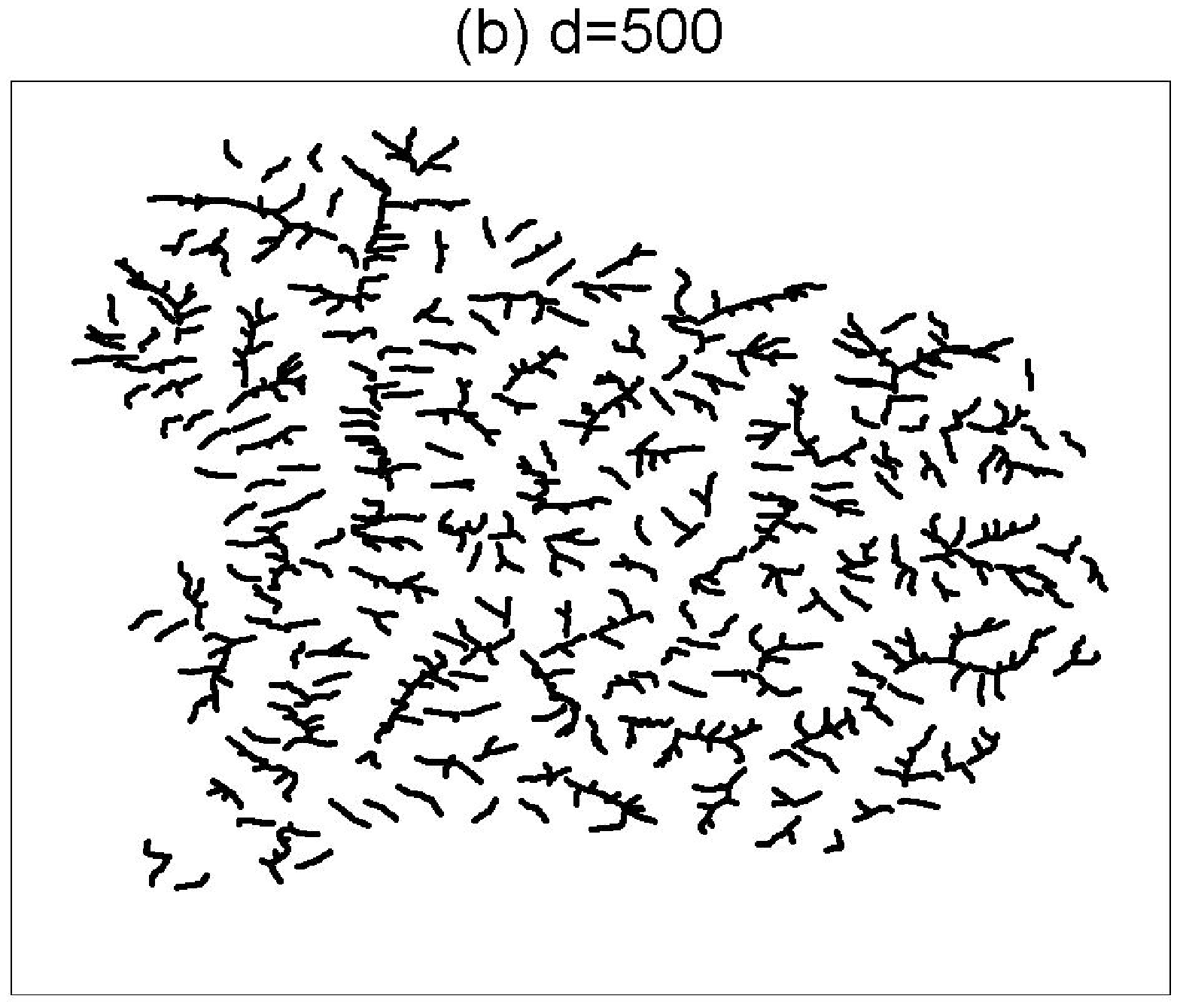}
\centering\includegraphics[width=.4\textwidth]{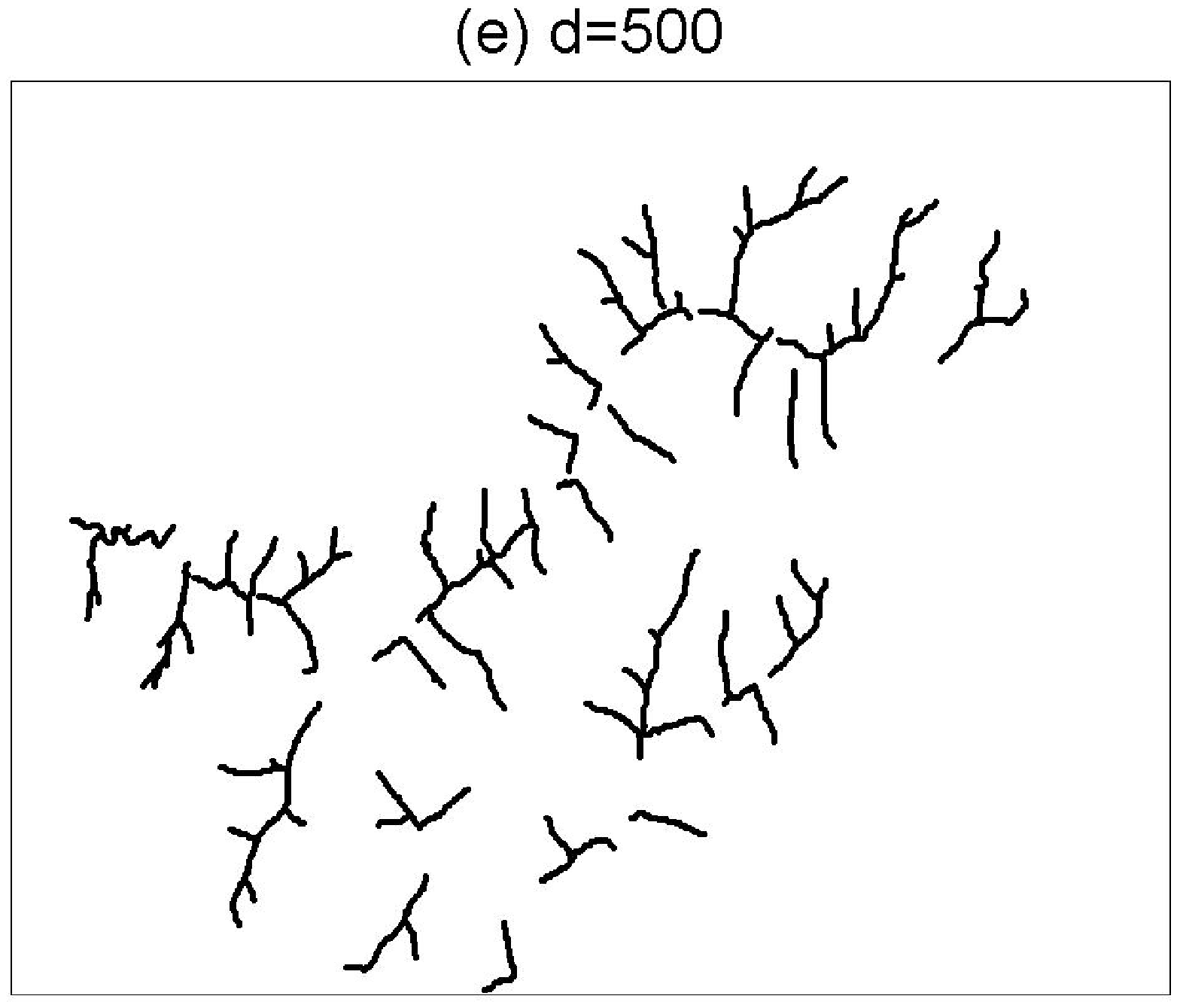}
\centering\includegraphics[width=.4\textwidth]{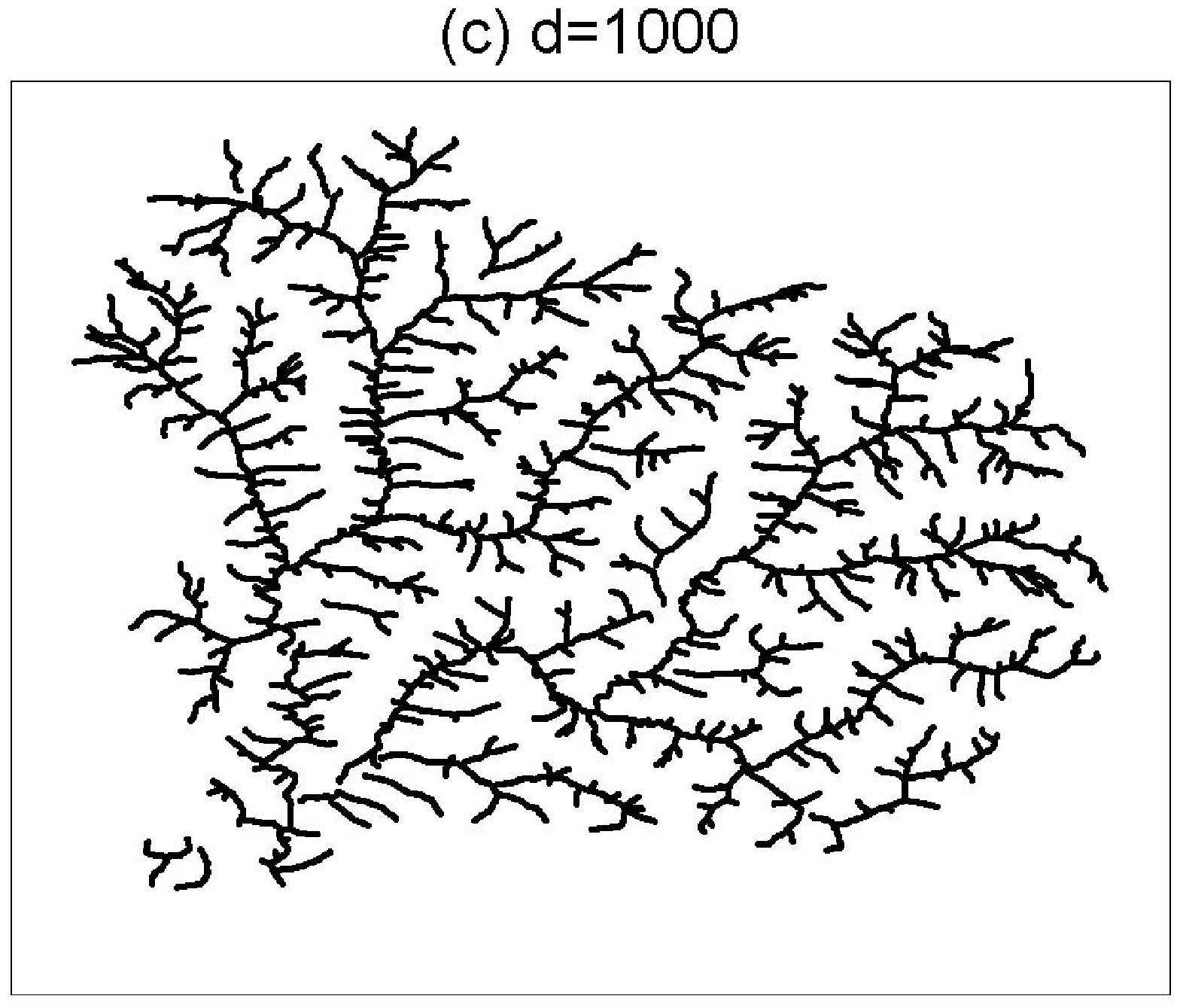}
\centering\includegraphics[width=.4\textwidth]{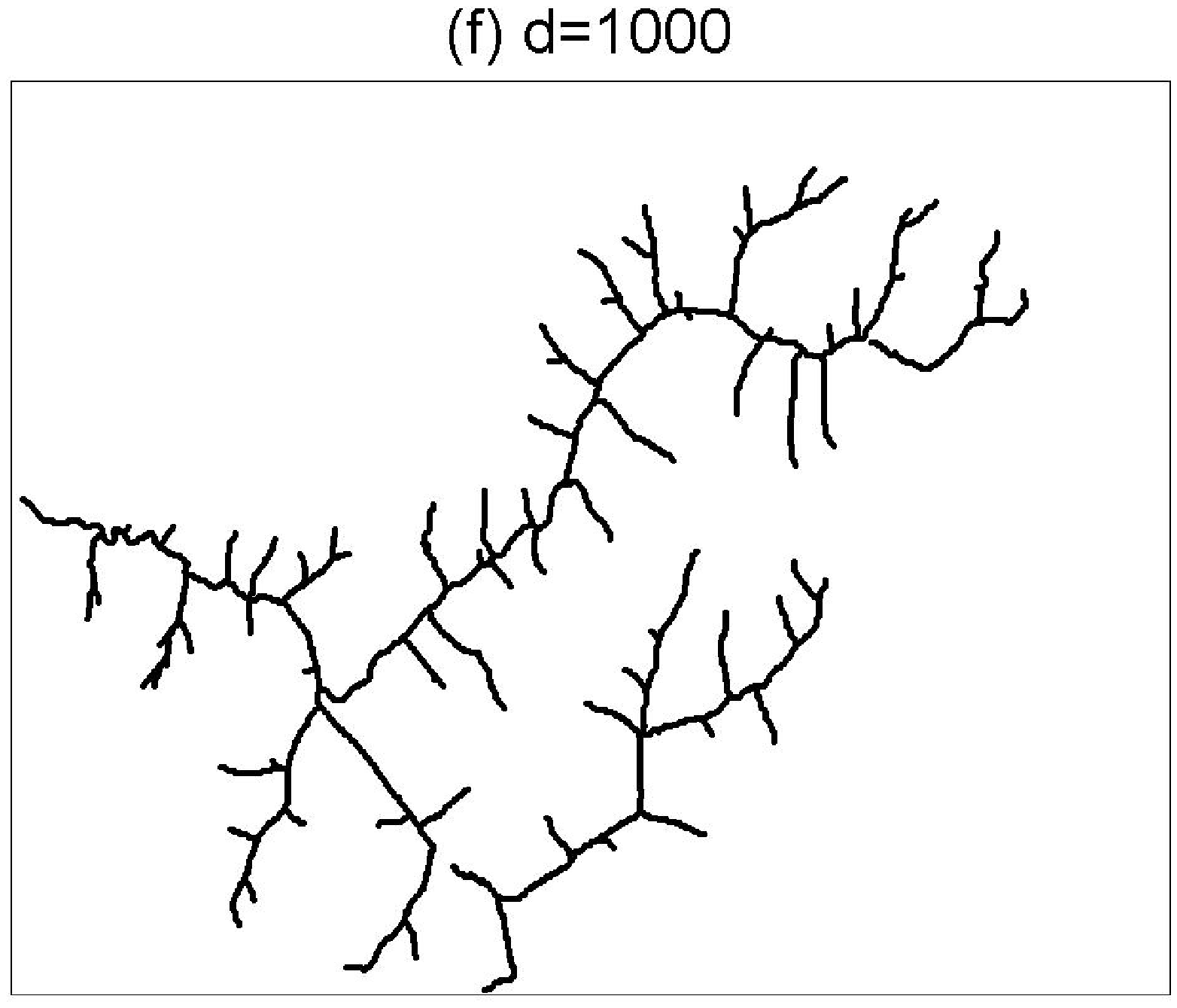}
\caption{Transport down the Noyo stream network.
Three snapshots of flux propagation from the stream sources to
the outlet, at (a,d) $d=200$, (b,e) $d=500$, and (c,f) $d=1000$.
Panels (a)--(c) show the entire Noyo basin, while
panels (d)--(f) zoom onto an order-4 subbasin located in the lower right
part of the entire basin.
See also Fig.~\ref{fig_snap0}.}
\label{fig_snap}
\end{figure}

\begin{figure}[p] %[p] [t]
\centering\includegraphics[width=.7\textwidth]{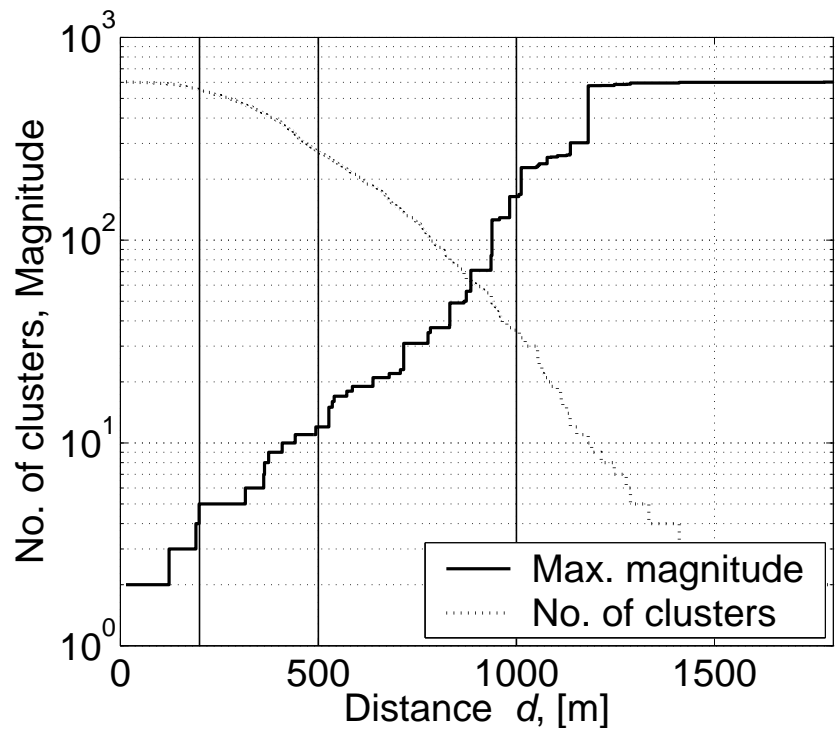}
\caption{Cluster evolution for the Noyo downstream flux transport:
number (dotted line) and largest-cluster size (solid line). Vertical
lines correspond to the three snapshots in Fig.~\ref{fig_snap}.}
\label{fig_snap0}
\end{figure}

This phase transition is further illustrated in Fig.~\ref{fig_snap}, which
shows three snapshots of the dye propagating down the Noyo basin.
The distances traveled by the dye at these snapshots are marked by
vertical lines in Fig.~\ref{fig_snap0}; the figure shows the number of clusters
(dotted line) and the magnitude of the largest cluster for the Noyo dynamic
tree (solid line), as a function of downstream propagation distance.

\begin{figure}[p] %[p] [t]
\centering\includegraphics[width=.7\textwidth]{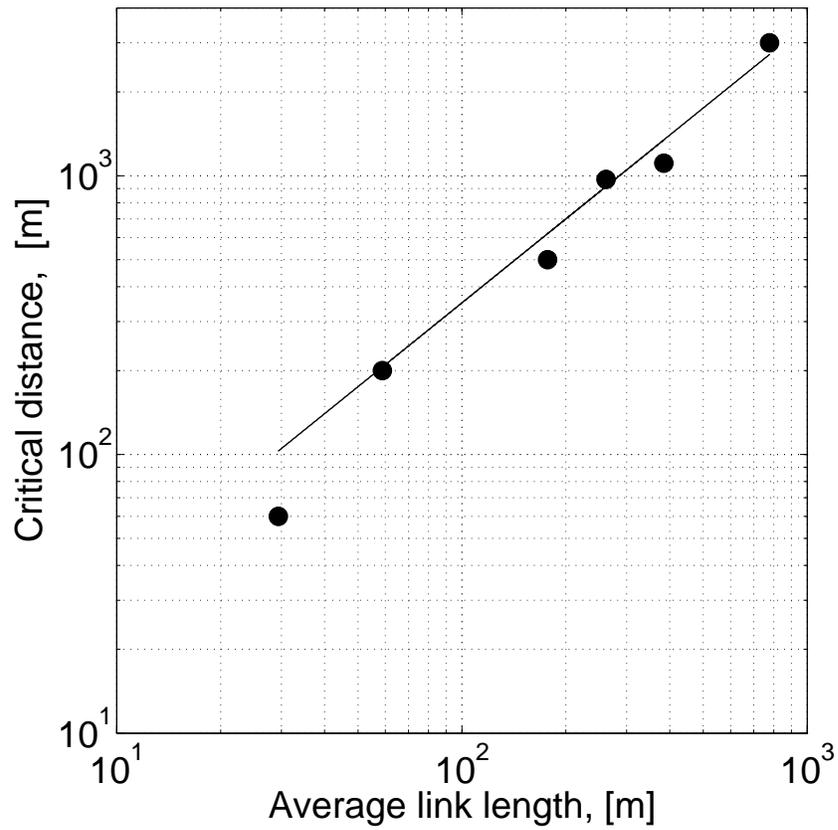}
\caption{Critical distance $d^*$ as a function of the average link
length $\bar{L}$ for the six dynamic trees shown in Fig.~\ref{fig_PT}.
The line in the figure corresponds to $d^*=3.5\,\bar{L}$.}
\label{fig_dc}
\end{figure}

The values of the six critical distances $d^*$ shown in Fig.~\ref{fig_PT}
vary over two orders of magnitude and depend strongly on the particular
network being analyzed.
Nevertheless, we notice a very good power-law fit for the
value of $d^*$ in terms of the average link length $\bar{L}$ of the
corresponding static tree (see Fig.~\ref{fig_dc}):
\be
d^*\approx 3.5\,\bar{L}.
\label{dL}
\ee
This relationship can be interpreted as follows in terms of the transport
on river networks: the giant cluster of connected streams is formed when each
flux traveled approximately 3.5 links downstream from a source.
We conjecture that: (a) this is a universal property of downstream transport on
Tokunaga trees with rich branching, {\it i.e.} Tokunaga SSTs with $c>1$ in Eq. \eqref{TSS};
(b) the coefficient of proportionality in \eqref{dL} may depend on the Tokunaga
parameters, but only weakly; and (c) this coefficient is larger than
or equal to 2 for {\it any} binary tree. An in-depth investigation of this issue
is left for future study.

\section{Concluding remarks}
\label{discussion}

\subsection{Summary}
\label{sum}

%% -->Efi, Ilya: At revision time we should add refs. & pointers to the figs.

This study focused on the statistical description of environmental transport
on self-similar river networks.
We approached the problem by considering downstream transport on such a network
as a particular case of nearest-neighbor {\it hierarchical aggregation}; the
so-called {\it ultrametric} induced by the branching structure of the river network
provides the distance function with respect to which the downstream flow
gives rise to clusters that decrease in number and increase in size with time.

We described the static topological structure of a drainage network
by the type of tree structure that goes back to the pioneering studies of
{\it Horton} (1945),
{\it Strahler} (1957) and
{\it Shreve} (1966),
and referred to it as a {\it static tree}, to
distinguish it from the associated {\it dynamic tree}. This novel concept introduced
herein describes downstream transport along the static tree.

We studied the statistical properties of both static and dynamic  trees using the
Horton-Strahler (HS) and Tokunaga branching taxonomies
[see {\it Horton,} 1945; {\it Strahler}, 1957; {\it Tokunaga}, 1978].
Using three river networks --- the Noyo,
Grigno and Tirso --- we showed that both static and dynamics trees can be well
approximated by Tokunaga {\it self-similar trees (SSTs)}. The HS
and Tokunaga parameters of these two types of trees differ
significantly, though, for each of the three basins.
This difference supports the relevance of the dynamic tree concept; its
parameter values depict important properties of the transport on a given river
network that are not captured by the conventional, static tree.

A striking result of this study is the phase transition we found in river network dynamics:
as one fills an empty river network through its sources,
or injects a dye into a water-filled one, the number of clusters of connected
nodes decreases and the size of the largest cluster increases, until
a dominant cluster of connected streams forms.
During this process, the time-dependent size distribution of the connected
clusters changes from an exponential to a power-law function as the critical time approaches.

This phenomenon, which may seem rather unexpected in the present, hydrological setting,
can be better understood within the framework of complex networks. This framework
has been explored in many natural and socio-economic settings, ranging from
the functioning of a cell to the organization of the internet 
[{\it Albert and Barabasi}, 2002].

The mathematical theory of complex networks considers a group of nodes that can
be connected with each other according to some problem-specific rules,
thus forming a graph.
In the simplest case, the node connections are independent of each other and can
be specified by the probability $p$ that two randomly chosen nodes are connected.
There exists a critical value $p_c$ such that for $p<p_c$
the network consists of isolated clusters, while a single giant
cluster appears as $p$ crosses $p_c$, and spans the entire network.
The appearance of this giant cluster is remarkably reminiscent of
infinite-cluster formation in percolation theory
[{\it Stauffer and Aharony}, 1994]. 
{\it Albert and Barabasi} [2002] provide
a comprehensive review of parallels and differences between complex-network
theory and percolation theory.

It readily follows from the analysis of Section~\ref{DT} that the transport on river
basins fits rather naturally the complex-network paradigm.
Formally, each river source is represented by a node and two streams are considered
to be connected when their respective fluxes join downstream.
This is exactly the scheme we used to define a dynamic tree,
with the only difference that we ignored the node connections within already
formed clusters.
This difference does not affect the process of cluster formation, so
all the results of complex-network theory do apply to
the envirodynamics of river basins.

There is an important difference, though, between complex networks
in general and the dynamic trees considered in this study.
Our dynamic trees, unlike general networks, are time-oriented, {\it i.e.},
their nodes can be ordered according in ``time'' or with respect to a ``distance'' parameter.
The ultrametric distance along such trees satisfies a
stronger triangle inequality than ordinary distance (see Section ~\ref{gen}).
Spaces equipped with an ultrametric $u$, instead of a traditional distance $d$,
have therefore interesting properties [{\it e.g.} {\it Schikhof}, 2007].
%% Noteworthy, a set of points in a metric space with a traditional distance $d$
%% naturally forms an ultrametric tree according to a nearest-neighbor clustering
%% procedure described in Section ~\ref{HA}.
%% MG-->IZ: Makes no sense to me, at least not in a hurry, like now :-)
As shown in Section ~\ref{HA}, hierarchical aggregation via nearest-neighbor clustering
provides a common framework for many apparently different processes ---
{\it e.g.,} billiards, river transport, and percolation --- in the setting of ultrametric trees,
and thus may provide novel insights into these processes.

%% We notice that in various models mentioned above the distribution of cluster magnitudes
%% (or masses) at the critical time is a power law with index that varies from model
%% to model (and seems to be different for different river networks).

In percolation models, the cluster-size distribution at phase transition is
given by a power law, whose index is a function of the system's dimension alone.
%% is universal for systems with the same dimension.
In our three river networks, this index differs from the one to the other, and from
the river to the hillslope network for the same basin.
In our hierarchical aggregation on dynamic trees, different clustering rules
may correspond to different effective ``dimensions'' of the system.
At the same time, it is known that the critical percolation indices are universal
for systems in high dimensions [{\it Hara and Slade}, 1990] and trees are a
simple model for infinite-dimensional systems [{\it Albert and Barabasi}, 2002].
Thus, one expects to see the same values of the critical indices when working
with percolation on a tree.
From this perspective, the fact that our critical exponents vary from basin to basin,
and from river to hillslope trees, still needs to be understood.

\subsection{Discussion and further work}
\label{discuss}

In this study we considered only the simplest clustering rules for the river streams:
two streams belong to the same cluster if there is a connected path from one
stream to another along the river network.
This approach is patterned after percolation studies and allows for a
straightforward treatment.
It might however result in a situation when two streams belong to the same cluster
despite the fact that the respective fluxes are not mixed yet (think of two short
streams that merge with a spatially extended cluster at about the same time).
Formulating a physically more appropriate set of clustering rules might
yield more realistic results for a wealth of river networks with differing properties.

So far, we only investigated dynamic trees that have the same set of leaves as the
corresponding static tree; this corresponds to injecting a flux through the sources.
At the same time, it might happen that a flux of interest is injected into an
internal node, {\it e. g.}, an industrial pollutant from
a plant or nutrient production from a local biotic activity.
Such situations can be easily modeled by considering a dynamic tree whose set
of leaves samples the entire river network.

%% A good source of empirical constraints for the development of a
To construct a richer theoretical framework for transport on river networks one
may also model the transport along real and synthetic networks by using
Boolean delay equations (BDEs).
%% Pls. cite Dee & Ghil (1984), Ghil & Mullhaupt (1985).
%% provide a convenient modeling framework for
%% answering a wide range of specific questions about the network transport.
In BDEs, the discrete state variables describe the flux
through the river branches; naturally, the rules for updating these variables
inherit the child-parent relationship of the stream's static tree.
The parent variables are updated based on the values of the
children variables, after delays that correspond to time of flux propagation
from a child to its parent.
{\it Ghil et al.} [2008] reviewed recently BDEs and their applications to climate
and earthquake modeling. We expect such modeling to shed further light
on the complex and important problems of river transport.

%% BDE modeling of transport on river networks is left for a future study.

\section{Acknowledgements}
This research was supported by the National Center for Earth-surface Dynamics (NCED),
a Science and Technology Center funded by NSF under agreement EAR-0120914, as
well as by NSF grants EAR-0824084 and EAR-0835789.
We thank Paola Passalaqua for helping with the river network extraction.

\end{document}